\documentclass[aps,pra,twocolumn, nofootinbib, nobalancelastpage]{revtex4-2} 
\usepackage[english]{babel}
\usepackage[utf8]{inputenc}
\usepackage{graphicx}       
\usepackage{dcolumn}       
\usepackage{bm}            
\usepackage{amsfonts}
\usepackage{amsmath}
\usepackage{amssymb}
\usepackage{dsfont}
\usepackage{bbold}
\usepackage[colorlinks=true, linkcolor=blue, citecolor=blue, urlcolor=blue]{hyperref}
\usepackage[normalem]{ulem} 

\renewcommand{\vec}[1]{{\mathbf #1}}

\begin{document}

\title{Defect states in three-dimensional diamond photonic band gap crystals}

\author{Julia Rocha\(^1\)}
\email{julia.rocha@lpmmc.cnrs.fr}

\author{Bart A. van Tiggelen\(^1\)}
\email{deceased}

\author{Ad Lagendijk\(^{2,3}\)} 
\email{a.lagendijk@tue.nl}

\author{Willem L. Vos\(^{1,2,3}\)} 
\email{w.l.vos@tue.nl}

\author{Sergey E. Skipetrov\(^1\)}
\email{sergey.skipetrov@lpmmc.cnrs.fr}
\affiliation{\(^1\)Université Grenoble Alpes, Centre National de la Recherche Scientifique (CNRS),
Laboratoire de Physique et de Modélisation des Milieux Condensés (LPMMC), 38000 Grenoble, France \\
\(^2\)Complex Photonic Systems (COPS) group, Department of Science and Technology, 
University of Twente, P.O. Box 217, 7500 AE Enschede, The Netherlands\\
\(^3\)Complex Photonic Systems (COPS) group, Photonic and Semiconductor Nanostructures (PSN) Chair, Department of Applied Physics and Science Education (APSE), 
Eindhoven University of Technology, P.O. Box 513, 5600 MB Eindhoven, The Netherlands
}

\date{24 August 2026}


\begin{abstract}
We perform a theoretical study of defect states within the photonic band gap of three-dimensional diamond crystals composed of point scatterers and doped with substitutional defects. 
The defects introduce localized states inside the photonic band gap,
whose existence conditions and eigenfrequencies are expressed in terms of the on-site Green's function of the ideal defect-free crystal. 
Off-site Green's functions are also calculated as function of distance and are shown to vanish within approximately two unit cells. 
Finite-size effects are analyzed by comparing the results obtained in the infinite-crystal limit with numerical simulations based on the coupled-dipole method. 
The latter not only reproduce the eigenfrequencies of the defect states within the band gap, but also provide  
their lifetimes originating from the finite crystal size. 
The lifetimes of the defect states increase exponentially with crystal size, becoming very long for large crystals. 
In addition to defect states in the three-dimensional photonic band gap, the defects also give rise to strongly detuned states outside the gap, which decouple from the spectrum of the ideal defect-free crystal.
\end{abstract}

\maketitle


\section{\label{sec:Introduction}Introduction}

The control of wave propagation using tailored nanostructures is a central goal of modern nanophotonics, enabling new functionalities with applications ranging from optical sensing to integrated circuits~\cite{Goldsmith2005book, Poslad2009Book, Saunders2025book}. 
In particular, confining light at the nanoscale has attracted sustained interest~\cite{Joannopoulos2008book, Lourtioz2008book, Novotny2012book2, carminati2021principles}. 
A widely used strategy for confining light relies on resonances, including optical cavities, coupled-cavity systems, plasmonic structures, and bound states in the continuum~\cite{Vahala2003Nature, Barnes2003Nature, Ghulinyan2015book, Hsu2016NatureReview}. Many of these nanophotonic tools modify the density of states (DOS) on an underlying continuum of (vacuum) states. Consequently, the pursued photonic confinement functionality is unavoidably competing with background effects caused by this continuum.

A complementary route consists in first creating a frequency interval in which optical states are forbidden in the targeted volume. 
When this inhibition occurs for all propagation directions and polarizations, the system exhibits a three-dimensional (3D) photonic band gap. 
Such gaps are realized in engineered nanostructures, called photonic crystals: composite dielectric media where the dielectric function varies periodically on length scales $a$ comparable to the wavelength of the light $a \simeq \lambda$~\cite{Bykov1972JETP, yablonovitch1987inhibited, john1987strong, sakoda2005optical, Joannopoulos2008book, Lourtioz2008book, Economou2010book}. 
Selected optical states can be introduced inside a photonic band gap through suitable structural modifications in a photonic crystal, such as the presence of defects~\cite{yablonovitch1991photonic, Joannopoulos2008book}. This work focuses on this class of photonic control, with the aim of developing analytical descriptions that provide physical insight into defect-induced states within the gap of a 3D photonic crystal. 

The existence of photonic band gaps is of central interest in quantum optics, as the modulation of the local DOS allows one to strongly enhance or suppress the spontaneous emission rate of embedded emitters~\cite{lodahl2004controlling, leistikow2011inhibiteda}. 
Furthermore, the depletion of the DOS favors Anderson localization, making photonic band gap structures a natural platform for the study of Anderson localization of light~\cite{john1987strong, john1991localization}. 
Recently, a new regime of light transport, in which light propagates by hopping between coupled cavities inside a complete photonic band gap, has been observed \cite{hack2019cartesian, adhikary2024observation}. This new kind of transport, called ``Cartesian light'', highlights the emergence of new physical phenomena within photonic band gaps and suggests new opportunities for controlling light transport in 3D as well as for the study of localization phenomena. 

While several dielectric structures exhibit a complete photonic band gap \cite{ho1990existence, yablonovitch1991photonic, sozuer1992photonic}, crystals composed of point dipoles are of particular interest for the study of light scattering, as they are realized in arrays of ultracold atoms~\cite{bloch2005ultracold, anderlini2007controlled}. 
Among such systems, the diamond crystal is the simplest structure presenting an omnidirectional photonic band gap \cite{antezza2009fanohopfield}. 
Its band gap persists over a broad range of lattice constants, widening for denser systems and closing for sparser crystals \cite{antezza2009fanohopfield}. 
The gap is also robust against common experimental imperfections, including finite-size effects, weak positional disorder, and the presence of vacancies \cite{skipetrov2020finitesize, antezza2013photonic}. 
These perturbations do not close the band gap, but instead introduce additional spectral features: surface states appear within the photonic band gap in finite systems with their DOS scaling with the inverse of the crystal size~\cite{Hasan2018PRL, skipetrov2020finitesize}; the presence of vacancies give rise to spatially localized states with frequencies lying within the band gap \cite{antezza2013photonic}; and mobility edges emerge from weak positional disorder \cite{skipetrov2020localization}.

In this work, we investigate defect-induced states within the 3D photonic band gap of a diamond crystal made of resonant point scatterers. 
We consider a substitutional disorder model in which a lattice site is occupied by a defect scatterer with resonance frequency distinct from those of the host scatterers, while the crystal structure remains unchanged. 
While positional disorder is often an unavoidable consequence in experimental realizations, substitutional disorder provides a powerful tool for spectral engineering, enabling a photonic analogue to electronic doping in semiconductors. Furthermore, it serves as a photonic realization of diagonal disorder which forms a building block for the Anderson model in
condensed matter physics. 
We begin our analysis by computing the Green’s functions for an infinite defect-free crystal. 
By using these functions, we demonstrate that the frequency of defect-induced states emerging within the band gap is tuned via the defect scatterers' resonance frequency. These predictions are compared with those obtained for finite systems that allow us to analyze finite-size effects that could be present in an experiment. 
We show that the defect states are spatially localized around the defect and exhibit significantly lower decay rates than those of typical states in the defect-free crystal. 
Finally, we characterize how the decay rates of the defect states scale with crystal size. 


\section{\label{sec:Model}Model}

\begin{figure*}
    \centering
    \includegraphics[width=0.98\linewidth]{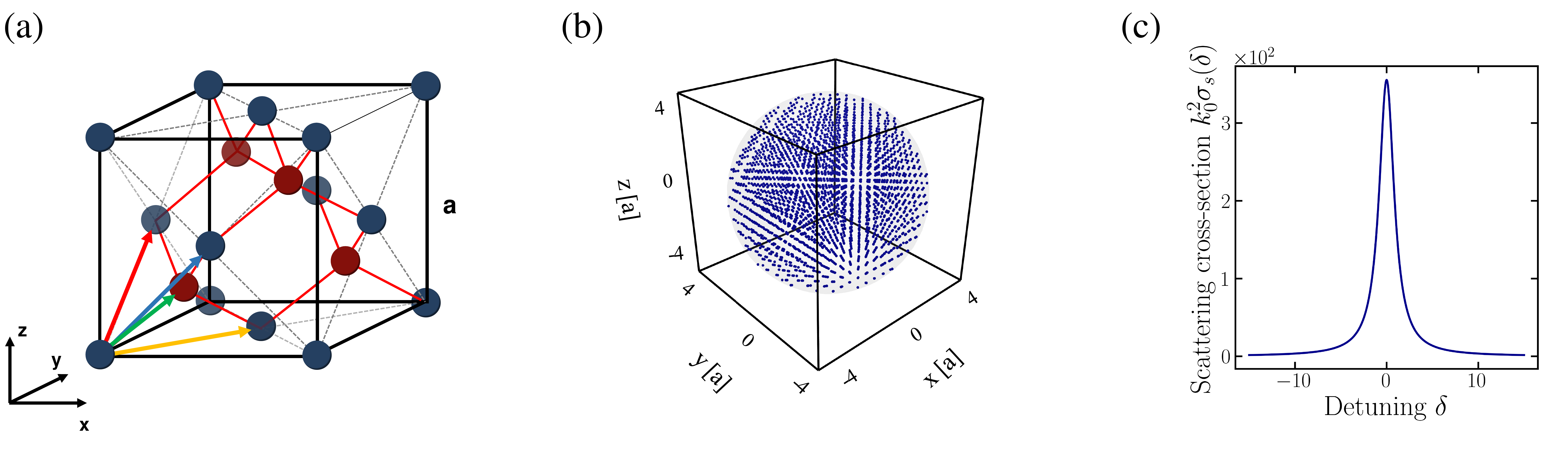}
    \caption{
    (a) Cubic unit cell of the diamond crystal structure. 
    Scatterers occupying sites \(A\) and \(B\) are shown as blue and red spheres, respectively, and their connecting vector \(\mathbf{d} = (a/4, a/4, a/4)\) is shown in green. 
    The primitive lattice vectors, \(\mathbf{a}_1 = (0, a/2, a/2)\), \(\mathbf{a}_2 = (a/2, 0, a/2)\) and \(\mathbf{a}_3 = (a/2, a/2, 0)\) are shown as red, yellow, and blue arrows, respectively. 
    (b) A crystal of finite size \(k_0R=15\) and lattice spacing \(k_0a=3.4\), made of \(N=2869\) scatterers. 
    (c) Dimensionless scattering cross-section of a resonant scatterer with polarizability $\alpha(\omega)$ defined by Eq.~\eqref{eq:polarizability1}.
    }
    \label{fig:DiamondLattice} 
\end{figure*}


\subsection{\label{subsec:FiniteLattices}Crystal of finite size}

We study the physical problem of light propagating in an ensemble of identical resonant scatterers arranged on the sites of a diamond crystalline structure. 
The latter is a non-Bravais structure with a basis of two sites per unit cell that can be viewed as two identical interpenetrating face-centered cubic (fcc) crystals with lattice constant \(a\)~\cite{ashcroft1976solid}. 
The sites of the first sublattice are generated by the primitive vectors \(\mathbf{a}_1 = (0, a/2, a/2)\), \(\mathbf{a}_2 = (a/2, 0, a/2)\), and \(\mathbf{a}_3 = (a/2, a/2, 0)\), and the second sublattice is obtained by translating the first by the displacement vector \(\mathbf{d} = (a/4, a/4, a/4)\). 
The sites of the first and second sublattices in the unit cell are denoted $A$ and $B$, respectively, as shown in Fig.~\ref{fig:DiamondLattice}(a).

A crystal of finite size is defined by taking the sites within a sphere of radius \(R\), unless explicitly stated otherwise, resulting in a total of \(N\) scatterers, see Fig.\ \ref{fig:DiamondLattice}(b). 
The propagation of electromagnetic waves at a frequency \(\omega\) in such a medium is governed by the Maxwell-Helmholtz equation~\cite{Joannopoulos2008book,  carminati2021principles, dalnegro2022}
\begin{equation}
    \nabla \times \nabla \times \mathbf{E}(\mathbf{r}) = \epsilon(\mathbf{r}) \left(\frac{\omega}{c}\right)^2 \mathbf{E}(\mathbf{r}),
    \label{eq:WaveEquation}
\end{equation}
where the permittivity \(\epsilon(\mathbf{r})\) is equal to
\begin{equation}
    \epsilon(\mathbf{r}) = 1 + \alpha_{\text{B}}\sum_{n=1}^{N} \delta (\mathbf{r} - \mathbf{r}_n),
    \label{eq:Epsilon}
\end{equation}
with \(\alpha_{\text{B}}\) the bare polarizability of a single scatterer, and \(\mathbf{r}_n\) the position of the \(n^\text{th}\) scatterer. 

The electric field \(\mathbf{E}_m\) that excites the scatterer on site \(\mathbf{r}_m\) is a superposition of the incident field \(\mathbf{E}_0(\mathbf{r}_m)\) and the fields scattered by all other scatterers at positions \(\mathbf{r}_n \ne \mathbf{r}_m\) \cite{carminati2021principles,dalnegro2022}
\begin{equation}
    \mathbf{E}_m = \mathbf{E}_0(\mathbf{r}_m) - k^2\alpha(\omega)\sum_{n\neq m}^N\mathcal{G}_0(\mathbf{r}_m-\mathbf{r}_n, \omega)\mathbf{E}_n.
    \label{eq:coupled}
\end{equation}
In this expression 
\begin{equation}
    \mathcal{G}_0(\mathbf{r},\omega) = -\frac{e^{i k r}}{4\pi r} \left[P(ikr)\mathds{1} + Q(ikr)\frac{\mathbf{r}\otimes\mathbf{r}}{r^2}\right]
    + \frac{\delta(\vec{r})}{3 k^2} \mathds{1} 
    \label{eq:DyadicGreen}
\end{equation} 
is the dyadic Green's function for electromagnetic waves, with
\(P(z) = 1 - 1/z + 1/z^2\), \(Q(z) = -1 + 3/z - 3/z^2\), the wave vector \(k = \omega/c\), and
\begin{equation}
    \alpha(\omega) = -\alpha(0) \frac{\omega_0^2}{\omega^2-\omega_0^2 +i \omega^3 \Gamma_0/\omega_0^2}
    \label{eq:polarizability}
\end{equation}
the dynamic polarizability of a single point scatterer with squared resonance frequency \(\omega_0^2 = 6\pi c^2/\Lambda_\mathrm{T} \alpha(0)\), zero-frequency polarizability \(\alpha(0) = \alpha_\mathrm{B}/(1 + \alpha_\mathrm{B} \Lambda_\mathrm{L}^3/6\pi)\), and bandwidth \(\Gamma_0 = \omega_0^2/c\Lambda_\mathrm{T}\). 
The inverse of the bandwidth $1/\Gamma_0$ gives the lifetime of the resonant state of the scatterer.
\(\Lambda_{\text{T}}^{-1}\) and \(\Lambda_{\text{L}}^{-1}\) are microscopic cut-offs lengths that regularize the divergence of \(\mathcal{G}_0(\mathbf{r},\omega)\) for \(r \to 0\) \cite{lagendijk96, vries1998point}\footnote{Note that Eq.\ \eqref{eq:polarizability} coincides with the polarizability of a two-level atom that is derived by considering the coupling of the atom with the electromagnetic vacuum without resorting to the model \eqref{eq:Epsilon} and the subsequent introduction of cut-off lengths $\Lambda_\mathrm{T}^{-1}$ and $\Lambda_\mathrm{L}^{-1}$ \cite{cohen98}. However, the microscopic approach still requires to deal with the divergence of $\mathcal{G}_0(\vec{r},\omega)$, whose divergent contribution is absorbed into a shift of the scatterer resonance frequency \(\omega_0\).}.

Equations~\eqref{eq:coupled} were originally derived by Foldy and by Lax in the context of the multiple scattering of scalar waves~\cite{foldy1945multiple, lax1951multiple}. 
As noted later on by Rusek \textit{et al.}~\cite{rusek1995analytical, rusek2000random}, it is convenient to consider quasinormal modes of these equations, defined as solutions in the absence of the incident field $\vec{E}_0(\vec{r})$. 
Such solutions may only exist at frequencies $\omega$ that obey the condition
\begin{equation}
\operatorname{det}\left[\mathds{1} + k^2\alpha(\omega)\mathds{G}_0(\omega) \right] = 0,
    \label{eq:resonanceG_0}
\end{equation}
where the matrix \(\mathds{G}_0 \) is composed of \(N\times N\) blocks, each of size \(3\times3\), given by the dyadic Green's function between a pair of scatterers in the crystal, whereas the diagonal blocks are zero. 
The elements of the block \(mn\) are 
\begin{equation}
\left[\mathds{G}_0(\omega)\right]_{mn}^{\mu\nu} = (1 - \delta_{mn})
    {\cal G}_0^{\mu\nu}(\vec{r}_{m}-\vec{r}_n, \omega),
    \label{eq:GMatrixElements}
\end{equation}
where \(\mu, \nu = x,y,z\) denote the projections on the axes of the Cartesian coordinate system. 
Equation~\eqref{eq:resonanceG_0} captures all poles of the total scattering operator except those associated with free fields unaffected by matter, i.e., special field configurations that vanish at every scatterer in the crystal~\cite{klugkist2006mode, antezza2009fanohopfield}.

Solutions that satisfy Eq.~\eqref{eq:resonanceG_0} are collective complex resonances of the ensemble of scatterers. In general, determining these solutions is difficult because the resonance condition is nonlinear in \(\omega\), with both the polarizability and the Green's function depending on the frequency. We therefore restrict our analysis to a narrow frequency interval around the single-scatterer resonance, such that $|\omega-\omega_0| \ll \omega_0$. In this regime, the polarizability in Eq.~(\ref{eq:polarizability}) is well approximated by 
\begin{equation}
    \alpha(\omega) \simeq -\frac{6\pi}{k_0^3}\frac{\Gamma_0/2}{\omega-\omega_0 +i \Gamma_0/2}
    =
    -\frac{6\pi}{k_0^3}\frac{1}{\delta +i},
    \label{eq:polarizability1}
\end{equation}
where we define a dimensionless frequency detuning 
\begin{equation}
\delta \equiv \frac{2(\omega-\omega_0)}{\Gamma_0}. 
\label{eq:detuning}
\end{equation}
The scattering cross-section \(\sigma_s(\omega)=(k^4/6\pi)|\alpha(\omega)|^2\) of a point scatterer with the polarizability given by Eq.~\eqref{eq:polarizability1} is shown in Fig.\ \ref{fig:DiamondLattice}(c). 
It has the usual Lorentzian line shape centered at \(\delta=0\) with width \(\Gamma_0\). 

In addition to the near-resonance approximation, we replace $\mathds{G}_0(\omega)$ by $\mathds{G}_0(\omega_0)$ in Eq.\ \eqref{eq:resonanceG_0}. This replacement is justified if the phase variation corresponding to changing the wave number from \(k\) to \(k_0\) remains small on the scale of crystal size: $|k - k_0| \times 2R \ll 2\pi$, or 
equivalently \(|\omega - \omega_0| \ll \pi c/R\).
In the vicinity of the single scatterer resonance or, more precisely, for detunings $|\delta| < \delta_{\text{max}}$, we arrive at the condition $\delta_{\text{max}} \ll 2\pi Q/k_0 R \simeq 10^6$, for a typical atomic transition in the optical range (quality factor $Q = \omega_0/\Gamma_0 \simeq 10^7$) and the largest length scale $k_0 R = 30$ that we analyze in this work. 
The detunings considered below are much smaller than this bound, so the approximation \(\mathds{G}_0(\omega)\simeq\mathds{G}_0(\omega_0)\) is well justified.
The resonance condition~\eqref{eq:resonanceG_0} can then be expressed in terms of an effective non-Hermitian Hamiltonian
\begin{equation}
        \mathcal{H} = \left(\omega_0 - i\frac{\Gamma_0}{2}\right)\mathds{1} - \frac{\Gamma_0}{2}\tilde{\mathds{G}}_0(\omega_0)
        \label{eq:effective_hamiltonian}
\end{equation}
as
\begin{equation}
\operatorname{det}\left[\left(\omega_l - i\frac{\Gamma_l}{2}\right)\mathds{1}-\mathcal{H} \right]=0,
        \label{eq:ResonanceCondition}
\end{equation}
where \(\tilde{\mathds{G}}_0(\omega) = -(6\pi/k) \mathds{G}_0(\omega) \).
Note that $\omega$ denotes the real part of the resonance from here on, whereas $-\Gamma/2$ corresponds to its imaginary part. 
According to Eqs.\ (\ref{eq:effective_hamiltonian}) and (\ref{eq:ResonanceCondition}), the complex resonances \(\omega_l - i\Gamma_l/2\) are related to the eigenvalues \(\Lambda_l\) of the matrix \(\tilde{\mathds{G}}_0(\omega_0)\) via
\begin{equation}
    \left\{
    \begin{aligned}
        \omega_l &= \omega_0 - \frac{\Gamma_0}{2}\operatorname{Re}\Lambda_l,\\
        \Gamma_l &= \Gamma_0 \left(1+\operatorname{Im}\Lambda_l\right).
    \end{aligned}
    \right.
    \label{eq:ResonancesFiniteCrystal}
\end{equation}
The DOS \(\mathcal{N}_N\left(\omega\right)\) for a crystal of finite size is computed from the eigenvalues of the effective Hamiltonian (\ref{eq:effective_hamiltonian}) as \cite{skipetrov2020finitesize}
\begin{equation}
    \mathcal{N}_N \left(\omega\right) = \frac{1}{3\pi N}\sum_{l=1}^{3N}\frac{\left(\Gamma_l/2\right)}{\left(\omega-\omega_l\right)^2 + \left(\Gamma_l/2\right)^2}\ ,
    \label{eq:DosFinite}
\end{equation}
with the normalization 
\begin{equation}
    \int_0^\infty d\omega\; \mathcal{N}_N(\omega) = 1.
    \label{eq:DosNorm}
\end{equation}

A right eigenvector \(\mathbf{\psi}_l\) of \(\tilde{\mathds{G}}_0(\omega_0)\) is a \(3N\)-dimensional vector that describes the spatial structure of the \(l\)-th quasinormal mode of the crystal. 
Its component \(\psi_l^{3(m-1)+\mu}\) is proportional to the \(\mu\)-polarization component
of the electric field on the scatterer at a
site \(\mathbf{r}_m\). The spatial localization of a state 
is characterized by the inverse participation ratio (\(\operatorname{IPR}\)) that is given by  
\begin{equation}
    \operatorname{IPR}_l = \sum_{m=1}^N\left\{\sum_{\mu=1}^3\left|\psi_l^{3(m-1)+\mu}\right|^2\right\}^2.
\end{equation}
The IPR quantifies how many scatterers sustain a quasinormal mode.
A quasinormal mode that is spatially localized on a single scatterer has \(\operatorname{IPR}_l = 1\) and a state extended over all $N$ scatterers in the crystal has \(\operatorname{IPR}_l = 1/N\). The matrix $\tilde{\mathds{G}}_0(\omega_0)$ also has left eigenvectors that we denote by $\phi_l$. 

\subsection{\label{subsec:InfiniteLattices}Infinite crystal}

In the infinite crystal limit, Bloch's theorem applies to Maxwell's equations~\cite{ashcroft1976solid, sakoda2005optical}. 
Thus, the fields that obey the Maxwell-Helmholtz equation \eqref{eq:WaveEquation} are expanded in Bloch modes \(\mathbf{E}_\mathbf{q}(\mathbf{r}) = \mathbf{u}_\mathbf{q}(\mathbf{r})e^{i\mathbf{q}\cdot\mathbf{r}}\), with \(\mathbf{q}\) a vector in the first Brillouin zone (BZ), and \(\mathbf{u}_\mathbf{q}(\mathbf{r})\) Bloch wave functions with the periodicity of the fcc crystal that underlies the diamond structure~\cite{antezza2009fanohopfield}. 
In this limit, the effective non-Hermitian \(3N\times3N\) Hamiltonian introduced in Eq.\ \eqref{eq:effective_hamiltonian} reduces to a \(6\times6\) Hermitian matrix of the form 
\begin{equation}
    \mathcal{H}(\mathbf{q}) = \left(\omega_0 - i\frac{\Gamma_0}{2}\right)\mathds{1} - \frac{\Gamma_0}{2}\tilde{\mathds{G}}(\mathbf{q}),
\label{eq:EffectiveHamiltonianInfinite}
\end{equation}
with the Green's matrix 
\begin{equation}
    \tilde{\mathds{G}}(\mathbf{q}) =
    \begin{bmatrix}
     \tilde{\mathds{G}}_{AA}(\mathbf{q}) &  \tilde{\mathds{G}}_{AB}(\mathbf{q}) \\
     \tilde{\mathds{G}}_{BA}(\mathbf{q}) &  \tilde{\mathds{G}}_{BB}(\mathbf{q})
    \end{bmatrix},
    \label{eq:GMatrixInfinite}
\end{equation}
where the \(3\times3\) blocks \(\tilde{\mathds{G}}_{\alpha\beta}\) are given by the sum over the reciprocal lattice
\begin{equation}
\begin{aligned} \tilde{\mathds{G}}_{\alpha\beta}(\mathbf{q}) &= -\frac{6\pi}{k} \frac{1}{\Omega} \sum_m \hat{\mathcal{G}}_0\left( \mathbf{b}_m - \mathbf{q}, \omega_0 \right) e^{i\varepsilon_{\alpha\beta}(\mathbf{b}_m - \mathbf{q}) \cdot \mathbf{d}}
    \\
    &+ \frac{6\pi}{k} 
    \delta_{\alpha\beta}
    \mathcal{G}_0(\mathbf{r} = 0, \omega_0),
\end{aligned}
\label{eq:G_alpha_beta}
\end{equation}
where \(\alpha = A,B\) denote the two scatterers in the unit cell, \(\varepsilon_{\alpha\beta}\) is the Levi-Civita symbol, 
\begin{equation}
\hat{\mathcal{G}}_0(\mathbf{q}, \omega_0) = \frac{\left(\vec{q} \otimes \vec{q}\right)/q^2}{k_0^2} + \frac{\mathbb{1} - \left(\vec{q} \otimes \vec{q}\right)/q^2}{k_0^2 - q^2 + i0^+}
\label{g0q}
\end{equation}
is the Fourier transform of the dyadic Green's function $\mathcal{G}_0(\mathbf{r}, \omega_0)$ defined by Eq.\ \eqref{eq:DyadicGreen}, and
$0^+$ denotes an infinitesimal positive real number. 
The reciprocal lattice vectors \(\mathbf{b}_m\) are written as \(\mathbf{b}_m = m_1\mathbf{b}_1 + m_2\mathbf{b}_2 + m_3\mathbf{b}_3\), with \(m_i \in \mathds{Z}\) and reciprocal lattice vector basis \(\{\mathbf{b}_j\}\) satisfying \(\mathbf{a}_i\cdot\mathbf{b}_j = 2\pi\delta_{ij}
\). The unit cell volume \(\Omega\) is computed from the primitive vectors of the direct lattice as \(\Omega = \mathbf{a}_1\cdot (\mathbf{a}_2 \times \mathbf{a}_3) = a^3/4\). For a complete derivation of Eqs.~\eqref{eq:EffectiveHamiltonianInfinite}--\eqref{eq:G_alpha_beta}, see Appendix \ref{ap:hamiltonian}.

While both terms in Eq.\ \eqref{eq:G_alpha_beta} diverge for $\alpha = \beta$, it is remarkable that their difference does not \cite{antezza2009fanohopfield, antezza2009spectrum, perczel2017photonic}. In practice, to evaluate  Eq.\ \eqref{eq:G_alpha_beta} for $\alpha = \beta$, dealing with finite numbers only, we regularize the two divergencies by replacing 
$\hat{\mathcal{G}}_0\left(\mathbf{q}, \omega_0 \right)$ with $\hat{\mathcal{G}}_0(\mathbf{q}, \omega_0) \exp(-\eta^2 q^2)$, leading to 
\begin{equation}
\begin{aligned}
\mathcal{G}_0(\vec{r}=0, \omega_0) &\rightarrow 
\int \frac{d^3 \vec{q}}{(2\pi)^3} \hat{\mathcal{G}}_0(\vec{q}, \omega_0) e^{-\eta^2 q^2}
\\
&=
\mathds{1}\frac{k_0}{6\pi} \left\{ \frac{1}{\sqrt{\pi} \eta k_0} \left[ \frac{1}{4(\eta k_0)^2} - 1 \right]
\right.
\\
&+\left. 
 e^{-\eta^2 k_0^2} \left[\mathrm{Erfi}(\eta k_0) -i \right] \right\},
\end{aligned}
\label{greg}
\end{equation}
and then take the limit $\eta \to 0^+$.

Solving the eigenproblem $\mathcal{H}(\vec{q}) \psi(\vec{q}) = \omega(\vec{q}) \psi(\vec{q})$ yields six bands \(\omega_s(\mathbf{q})\), labeled \(s=1\)--\(6\).
As in the case of a crystal of finite size, they are related to the eigenvalues \(\Lambda_s(\mathbf{q})\) of the Green's matrix \eqref{eq:GMatrixInfinite}  via \(\omega_s(\mathbf{q}) = \omega_0 - \frac{\Gamma_0}{2}\operatorname{Re}\Lambda_s(\mathbf{q})\). 
As the crystal considered in this section is infinite, energy cannot escape and thus the decay rates of the modes vanish, yielding \(\Gamma_s(\mathbf{q}) = \Gamma_0 \left[1+\operatorname{Im}\Lambda_s(\mathbf{q})\right] = 0\). 
The normalized DOS $\mathcal{N}(\omega)$ in the infinite crystal is obtained by summing over the six bands and integrating over the first Brillouin zone, resulting in 
\begin{equation}
    \mathcal{N}(\omega) = \Omega
    \sum_{s=1}^6\int_{\text{BZ}} \frac{d^3\mathbf{q}}{(2\pi)^3}\delta\left[\omega-\omega_s(\mathbf{q})\right],
    \label{eq:DosInfinite}
\end{equation}
which is consistent with the limit \(\Gamma_l \rightarrow 0\) in Eq.\ \eqref{eq:DosFinite}.
We indeed verified that the DOS obtained from Eq.~\eqref{eq:DosFinite} converges to the DOS of the infinite diamond crystal given by Eq.~\eqref{eq:DosInfinite} as the number of scatterers $N$ increases.


\begin{figure}[t]
    \centering
    \includegraphics[width=\linewidth]{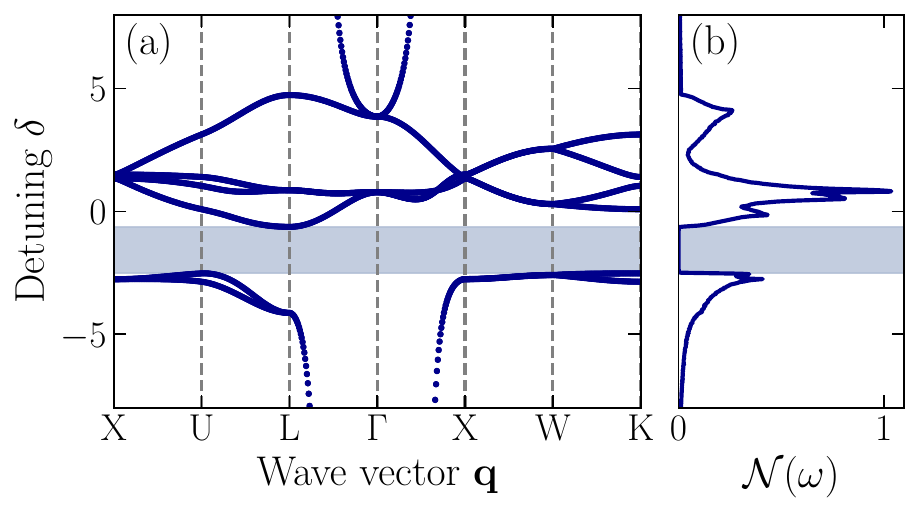}
    \caption{(a) Band diagram and (b) normalized DOS for light in a diamond crystal consisting of point scatterers with a lattice constant \(k_0a=3.4\). 
    The band diagram is computed along the standard irreducible path X–U–L–\(\Gamma\)–X–W–K in the first Brillouin zone of the underlying fcc crystal. 
    The DOS is obtained by uniformly sampling \(10^6\) \(\mathbf{q}\) vectors throughout the first Brillouin zone. 
    The shaded region indicates the complete photonic band gap.}
    \label{fig:BandDiagram}
\end{figure}

\section{ Optical modes and Green's functions for light in a defect-free diamond crystal} 

\subsection{Resonances of defect-free diamond crystals}
\label{sec:ResonancesInDefectFreeCrystals}

Figure~\ref{fig:BandDiagram}(a) shows the band diagram that is obtained by diagonalizing the matrix $\tilde{\mathds{G}}(\mathbf{q})$ given by Eq.\ \eqref{eq:GMatrixInfinite} for an infinite diamond crystal with lattice constant \(k_0a = 3.4\). 
The band diagram is computed by varying \(\mathbf{q}\) along the standard irreducible path in the first Brillouin zone of the crystal. 
For every wave vector \(\mathbf{q}\) we see 6 bands, as discussed above. 
The shaded frequency interval, \(\delta \in \left[-2.50, -0.62\right]\), contains no bands along the high-symmetry path shown in Fig.~\ref{fig:BandDiagram}(a). 

Figure~\ref{fig:BandDiagram}(b) shows the DOS that corresponds to the band structure in Fig.\ \ref{fig:BandDiagram}(a) obtained by numerically evaluating Eq.\ \eqref{eq:DosInfinite} as a sum over \(10^6\) \(\mathbf{q}\) vectors that are uniformly sampled across the entire first Brillouin zone of the crystal. 
The DOS extends over a detuning considerably larger than the range over which the scattering cross section of an isolated point scatter shown in Fig.~\ref{fig:DiamondLattice}(c) is appreciable. 
In the interval \(\delta  = 2(\omega-\omega_0)/\Gamma_0 \in \left[-2.50, -0.62\right]\), the DOS vanishes within numerical resolution, confirming that this frequency range corresponds to a complete photonic band gap. The band-edge frequencies obtained in this way are in good agreement with previous calculations for diamond photonic crystals of point scatterers~\cite{antezza2009fanohopfield}. These studies also showed that the band gap shrinks monotonically with increasing lattice constant, corresponding to decreasing scatterer number density, and closes for \(k_0a \gtrsim 5.14\)~\cite{antezza2009fanohopfield}.

\begin{figure}
    \centering
    \includegraphics[width=1\linewidth]{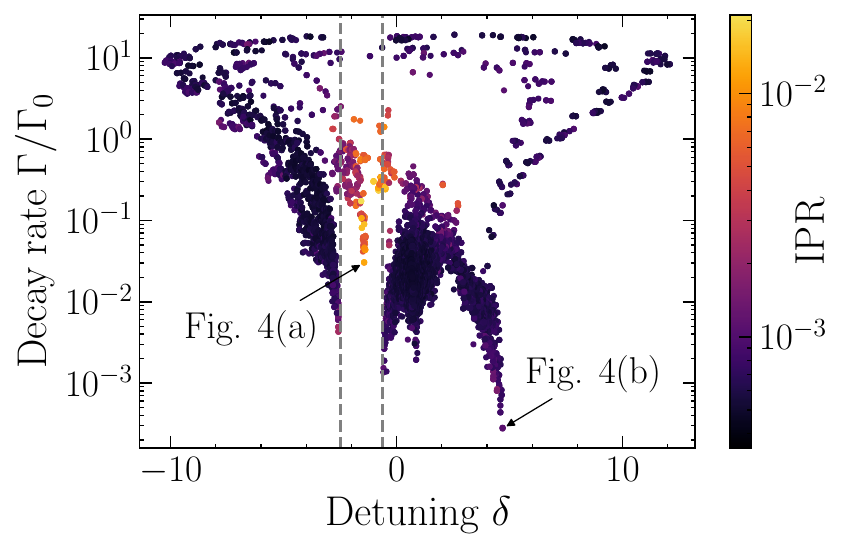}
    \caption{Distribution of resonances in the complex plane for a diamond crystal of lattice spacing \(k_0a = 3.4\) and radius \(k_0R=15\). Each eigenvalue is colored according to the IPR of the corresponding eigenvector, showing the degree of its spatial localization. Vertical dashed lines indicate the band edges of the infinite crystal.}
\label{fig:ResonancesDefectlessLattice}
\end{figure}

\begin{figure}
    \centering
    \includegraphics[width=\linewidth]{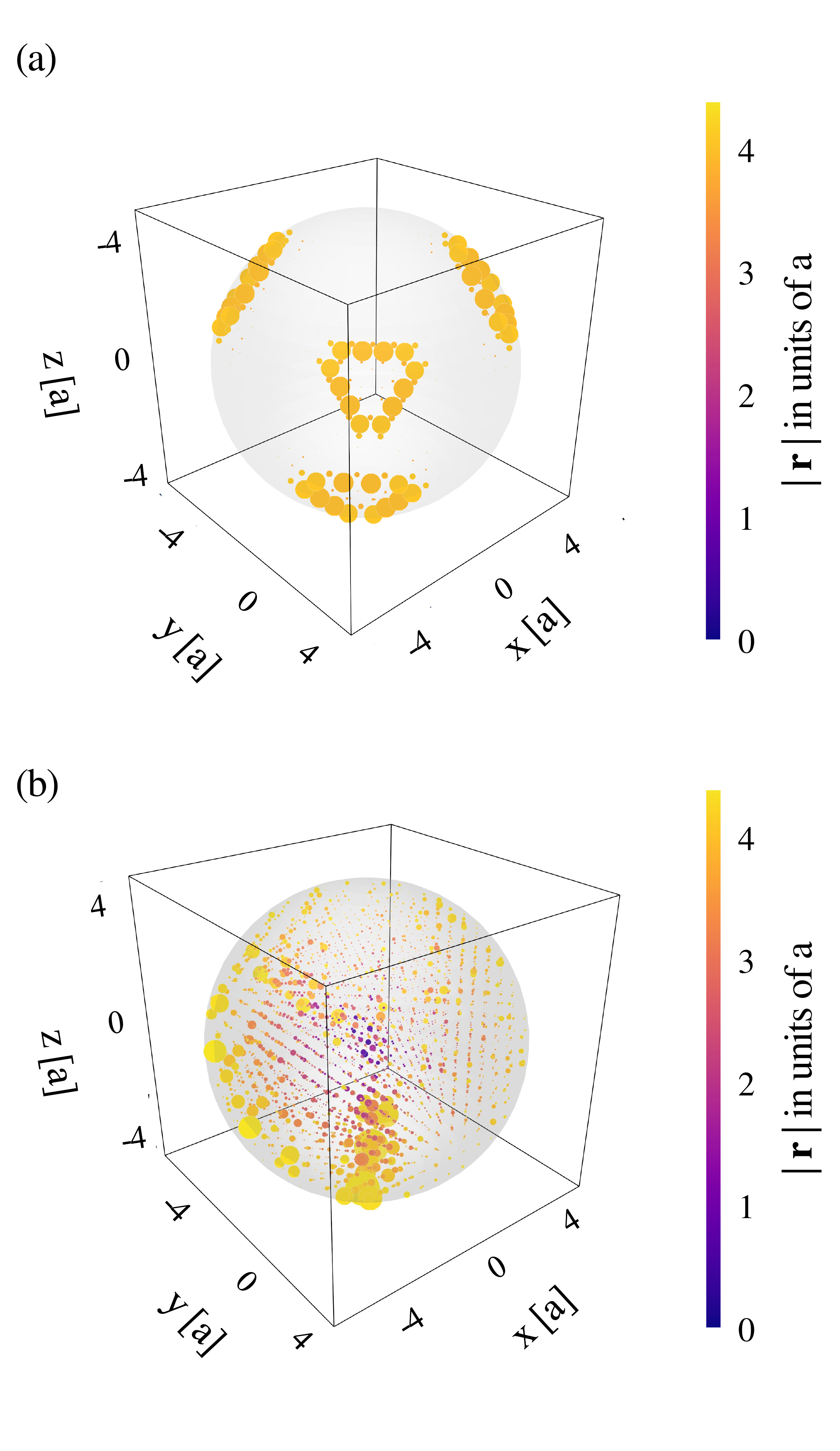}
    \caption{Spatial profiles of two representative states corresponding to the eigenvalues indicated by arrows in Fig.~\ref{fig:ResonancesDefectlessLattice}. 
    (a) Example of a surface state that has the longest lifetime among all states within the photonic band gap. 
    (b) Example of a bulk state localized inside the crystal with the longest lifetime outside the photonic band gap. 
    Each state is represented by \(N\) spheres centered at the lattice sites \(\{ \mathbf{r}_m \}\) and with radii proportional to the intensity
    \(I_l^m = \sum_{\mu=1}^3 |\psi_{l}^{3(m-1)+\mu}|^2\) of the state on the site $\vec{r}_m$. 
    The color scale encodes the depth within the sample, from deep violet for sites in the center of the crystal to yellow for sites at the crystal surface. 
    The gray sphere in both panels delineates the spatial extent of the finite crystal. } 
    \label{fig:DefectlessLatticeModes}
\end{figure}

For a finite crystal with radius \(k_0R=15\), corresponding to \(N = 2869\) point scatterers, and the same lattice constant \(k_0a=3.4\), the ensemble of resonances obtained from the effective Hamiltonian in Eq.~\eqref{eq:effective_hamiltonian} is shown in Fig.~\ref{fig:ResonancesDefectlessLattice}. 
The color scale represents the \(\operatorname{IPR}\) of the quasinormal modes associated with the eigenvalues shown in the figure. 
The majority of the states are extended and have \(\operatorname{IPR}\sim 1/N \sim 10^{-4}\). 
A small number of states arise within the band gap of the infinite crystal, delimited by vertical dashed lines in the figure. 
These states are confined to the surface of the sample with their normalized DOS scaling as the inverse of the sample radius \cite{skipetrov2020finitesize, Hasan2018PRL}. 
Due to this confinement, they feature slightly larger inverse participation ratios (IPR), reaching a maximum \(\operatorname{IPR} \simeq 0.021\). 
The spatial profile of the surface state indicated by an arrow in Fig.\ \ref{fig:ResonancesDefectlessLattice} is shown in  Fig.\ \ref{fig:DefectlessLatticeModes}(a). This state is the longest-lived state within the gap with a decay rate \(\Gamma/\Gamma_0 \simeq 3.0\times 10^{-2}\). It exhibits a well-defined symmetry, being confined to a small number of scatterers arranged on a triangular pattern that repeats itself four times at the surface of the crystal. 

\begin{figure}
    \centering
    \includegraphics[width=1\linewidth]{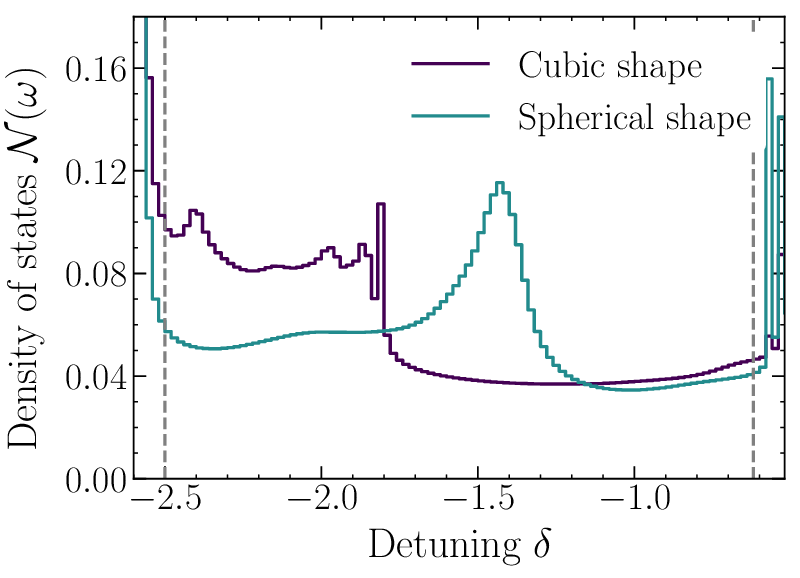}
    \caption{DOS of a diamond photonic crystal of finite size and lattice constant \(k_0a=3.4\), for two different sample shapes. Spherical (cubic) sample has radius \(k_0R=15\) (edge length \(k_0L=24\)) and contains \(N=2869\) (\(N=3059\)) scatterers.}
    \label{fig:DOSFinite}
\end{figure}

The distribution of surface states within the gap depends strongly on the size and shape of the finite crystal. 
We compare the DOS of the finite spherical crystal with the DOS of a finite cubic crystal with edge length \(k_0L = 12\) in Fig.~\ref{fig:DOSFinite}. 
The size of the cubic crystal is chosen to have a similar number of scatterers as in the spherical crystal. 
In the spherical crystal, the DOS exhibits a pronounced peak around \(\delta= -1.42\), whereas the behavior near the band edges is nearly symmetric, with two plateaus at \(\mathcal{N}(\omega) \simeq 0.055\) and \(\mathcal{N}(\omega) \simeq 0.040\) at the low and high-frequency band edges, respectively. 
In contrast, for the cubic crystal the central peak disappears and a significantly broader plateau is observed on the high frequency side of the gap, with \(\mathcal{N}(\omega) \simeq 0.04\) for \(\delta \in [-1.80, -0.62]\). 
When the detuning \(\delta\) decreases to \(\delta = -1.80\), the DOS increases rapidly and then fluctuates around \(\mathcal{N}(\omega)\simeq0.09
\) down to the low-frequency band edge. 
Differences between the DOS of spherical and cubic samples of diamond crystals were also reported for other lattice constants by Antezza and Castin~\cite{antezza2013photonic}.

Outside the band gap, on the high frequency side of the spectrum in Fig.~\ref{fig:ResonancesDefectlessLattice}, branches of states exhibiting low decay rates (long lifetimes) emerge.
Figure~\ref{fig:DefectlessLatticeModes}(b) shows the spatial structure of the longest-lived state indicated by an arrow in Fig.~\ref{fig:ResonancesDefectlessLattice}. 
This mode has decay rate \(\Gamma/\Gamma_0 = 2.7\times 10^{-4}\) and remains comparatively extended throughout the sample, demonstrating that the longest lifetimes do not necessarily arise from strong spatial localization. This highlights the distinction between spatial confinement and the suppression of radiative decay in finite photonic crystals.
The spatial distribution of the state is preserved in the cubic crystal, as is expected for a bulk state that is only weakly affected by sample boundaries. 
In the cubic crystal, however, this longest-lived state exhibits an even smaller decay rate \(\Gamma/\Gamma_0 = 2.6\times 10^{-5}\). 

\subsection{Green's function of diamond crystals}
\label{sec:GreensFunction}

It follows from the Maxwell-Helmholtz equation \eqref{eq:WaveEquation} and the expression for the permittivity \eqref{eq:Epsilon} that in the finite diamond crystal the Green's function \(\mathcal{G}_\text{C}(\mathbf{r},\mathbf{r}',\omega)\) satisfies 
\begin{equation}
\begin{aligned}
    \mathcal{G}_\text{C}(\mathbf{r},\mathbf{r}',\omega) &= \mathcal{G}_0(\mathbf{r}-\mathbf{r}',\omega) \\
    &- k^2\alpha_{\text{B}}\sum_{j = 1}^N\mathcal{G}_0(\mathbf{r}-\mathbf{r}_j,\omega)\mathcal{G}_\text{C}(\mathbf{r}_j,\mathbf{r}',\omega).
\label{eq:CoupledEquationsCrystalGreenFunction}    
\end{aligned}
\end{equation}
We now restrict our consideration to $\vec{r}$ and $\vec{r}'$ coinciding with lattice sites $\vec{r}_m$ and $\vec{r}_n$ and define the $3N \times 3N$ matrices $\operatorname{G}_{\text{C}}$ and $\operatorname{G}_0$ with elements $(\operatorname{G}_{\text{C}})_{mn} \equiv -(6\pi/k) \mathcal{G}_{\text{C}}(\vec{r}_m,\vec{r}_n,\omega)$ and $(\operatorname{G}_0)_{mn} \equiv -(6\pi/k) \mathcal{G}_0(\vec{r}_m-\vec{r}_n,\omega)$, for all \(m\) and \(n\).\footnote{Note that the matrix $\operatorname{G}_0$ is related to the Green's matrix \(\mathds{G}_0\) defined in \eqref{eq:GMatrixElements} by
\((\operatorname{G}_0)_{mn}= - \frac{6\pi}{k}\left[(\mathds{G}_0)_{mn}
+\delta_{mn}\mathcal{G}_0(0,\omega)\right]\).}
By expressing Eq.\ (\ref{eq:CoupledEquationsCrystalGreenFunction}) as a sum of scattering processes of increasing orders and summing the resulting infinite series, we can write $\operatorname{G}_{\text{C}}$ in terms of $\operatorname{G}_0$ as
\begin{equation}
    \operatorname{G}_{\text{C}} = \operatorname{G}_0-\frac{k}{6\pi} \operatorname{G}_0 \operatorname{T} \operatorname{G}_0, \label{eq:DysonEquation}
\end{equation}
with the $3N \times 3N$ scattering matrix $\operatorname{T}$ given by
\begin{equation}
    \operatorname{T} = -\frac{\mathds{1}}{\frac{1}{k^2\alpha(\omega)} +  \mathds{G}_0}\ .
    \label{hctmatrix}
\end{equation}

A lengthy but straightforward calculation, presented in Appendix \ref{ap:green}, shows that the eigenvectors $\psi_l$ of the matrix $\tilde{\mathds{G}}_0$, and hence eigenvectors of \(\mathds{G}_0\), are also eigenvectors of $\operatorname{G}_{\text{C}}$ with corresponding eigenvalues \(\zeta_l\) given by
\begin{equation}
    \zeta_l \simeq -Q_{\text{B}} \frac{Q_{\text{B}} + \Lambda_l + i}{\delta + \Lambda_l + i},
    \label{eq:EigenvaluesZeta}
\end{equation}
where $Q_{\text{B}} = 6\pi/k^3 \alpha_{\text{B}}$ and $|\omega-\omega_0| \ll \omega_0$ is assumed.

The matrix $\operatorname{G}_{\text{C}}$ admits a spectral decomposition $\operatorname{G}_{\text{C}} = \Psi \zeta \Phi^{\dagger}$, where \(\Psi\) and \(\Phi\) are \(3N\times3N\) matrices having the right and left eigenvectors, \(\psi_l\) and \(\phi_l\), as columns, respectively, and \({\zeta} = \operatorname{diag}\left[\zeta_1, \zeta_2,..., \zeta_{3N} \right]\) a \(3N\times3N\) diagonal matrix having the eigenvalues \(\zeta_l\) as elements. From this decomposition, the element $\mu\nu$ of the block $mn$ of the matrix $\operatorname{G}_{\text{C}}$ is given by
\begin{equation}
\begin{aligned}
    [\operatorname{G}_{\text{C}}(\delta)]_{mn}^{\mu\nu} &= - Q_{\text{B}} \sum_{l=1}^{3N}
    \frac{Q_{\text{B}} + \Lambda_l + i}{
    \delta + \Lambda_l + i}
    \\
    &\times
    \psi_{l}^{3(m-1)+\mu}
    \left[
    \phi_{l}^{3(n-1)+\nu}
    \right]^*,
    \end{aligned}
    \label{gcfinal}
\end{equation}
where $m$ and $n$ enumerate unit cells, and $\mu, \nu = x,y,z$ label the three polarization components.
Equation~\eqref{gcfinal} can be generalized straightforwardly to an infinite crystal. 
To this end, we replace the sum over discrete states $l$ by an integral over the first Brillouin zone and a sum over the six bands [see Fig.\ \ref{fig:BandDiagram}(a)]. Using the Hermiticity of the Hamiltonian in Eq.~\eqref{eq:EffectiveHamiltonianInfinite} and restricting ourselves to the regime \(|\delta| \ll Q_{\text{B}}\), we obtain
\begin{equation}
\begin{aligned}
&[\operatorname{G}_{\text{C}}(\delta)]_{mn, \alpha\beta}^{\mu\nu} =  -\Omega Q_{\text{B}}^2
    \\
    &\times\sum_{s=1}^{6}
    \int_{\text{BZ}} \frac{d^3 \vec{q}}{(2\pi)^3}
\frac{\psi^{\mu}_{s\alpha}(\vec{q})
    \left[
    \psi^{\nu}_{s\beta}(\vec{q})
    \right]^* e^{i \vec{q} \cdot (\vec{R}_m -\vec{R}_n)}}{
    \delta +\Lambda_s(\vec{q}) +i},
\end{aligned}
\label{gcfinalq}
\end{equation}
where $\vec{R}_m$ is the position of the midpoint between the scatterers $A$ and $B$ within the \(m^\text{th}\) unit cell, while $\alpha, \beta = A, B$ label the two scatterers within each unit cell.
Here $\psi^{\mu}_{s\alpha}(\vec{q})$ denotes the $\mu$-polarization component of the eigenvector $\psi_{s\alpha}(\vec{q})$ associated with the eigenvalue $\Lambda_s(\vec{q})$, evaluated on the scatterer $\alpha = A$ or $B$.

\begin{figure}[t!]
    \centering
    \includegraphics[width=\linewidth]{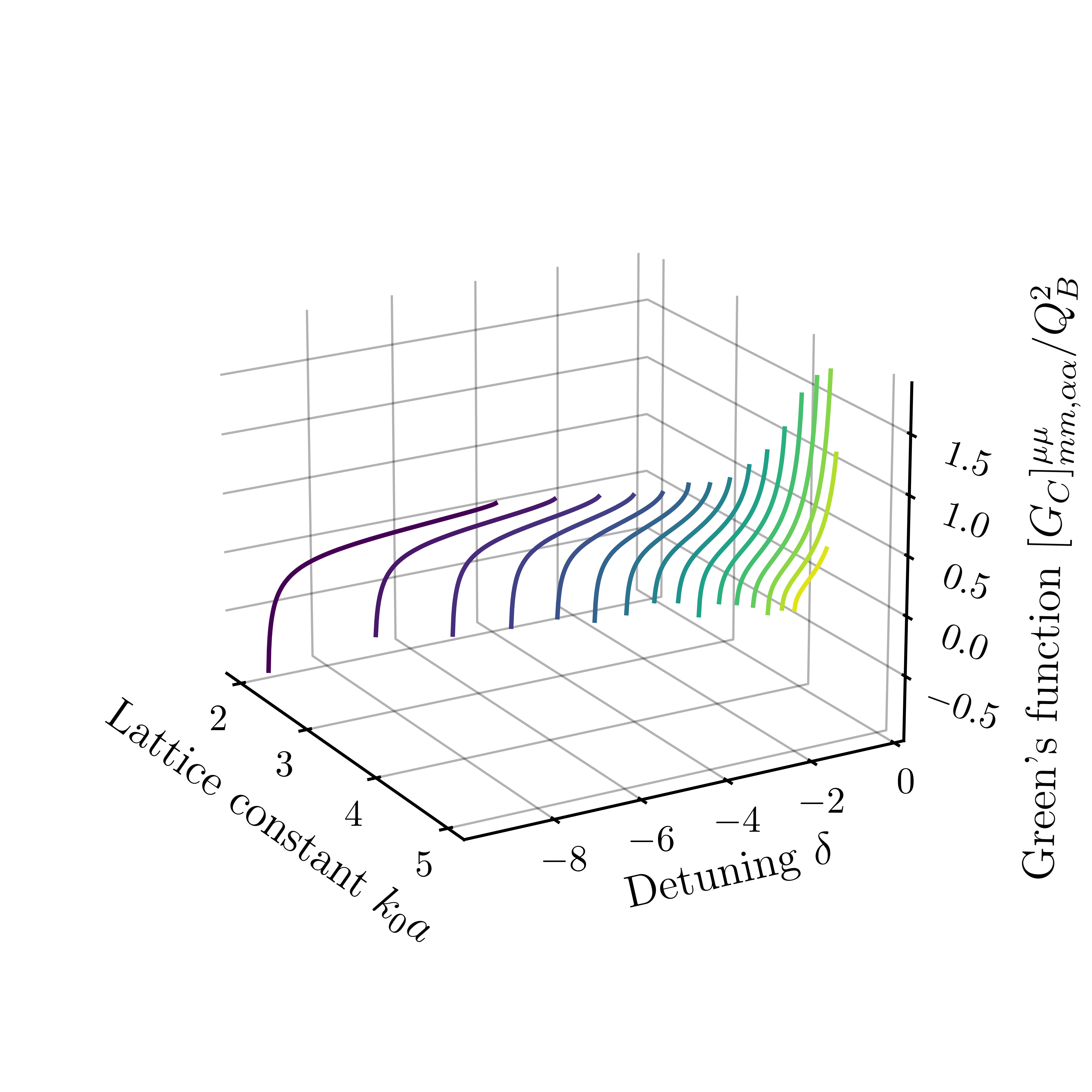}
    \caption{Green's function $[G_{\text{C}}]_{mm, \alpha\alpha}^{\mu\mu}$
    for a diamond crystal of point scatterers as a function of both the detuning \(\delta\) within the photonic band gap and the reduced lattice constant \(k_0a \in [2.0,5.0]\). 
    $[G_{\text{C}}]_{mm, \alpha\alpha}^{\mu\mu}$ is computed by numerically evaluating the integral in Eq. \eqref{gcfinalq} by uniformly sampling over \(10^6\) \(\mathbf{q}\) vectors in the first Brillouin zone.
    }  \label{fig:green_function}
\end{figure}

Figure~\ref{fig:green_function} shows the on-site Green's function of the crystal, corresponding to  \(m = n\), $\alpha = \beta$, and $\mu = \nu$, for lattice constants \(k_0a \in [2.0, 5.0]\) and detunings \(\delta\) within the band gap. Note that the band edges shift with $k_0 a$. The band-edge frequencies obtained in our calculations agree well with those reported in Ref.~\cite{antezza2009fanohopfield}.
The results are obtained by numerically evaluating the integral in Eq.\ \eqref{gcfinalq} over the entire Brillouin zone and, owing to the symmetries of the crystal, they are independent of $m$, $\alpha$, and $\mu$.
Within the photonic band gap, the Green's function $[G_{\text{C}}]_{mm, \alpha\alpha}^{\mu\mu}$ is real, since its imaginary part is proportional to the DOS, which vanishes inside the gap.
A clear asymmetry is observed between the behavior of the functions near the two band edges. 
Close to the high-frequency edge, the Green's functions remain positive for all values of lattice constants considered. 
In contrast, near the low-frequency edge, the Green's functions are negative for \(k_0a < 4.8\), and become positive for \(k_0a \geq 4.8\). 

\begin{figure*}[t]
    \centering
\includegraphics[width=\linewidth]{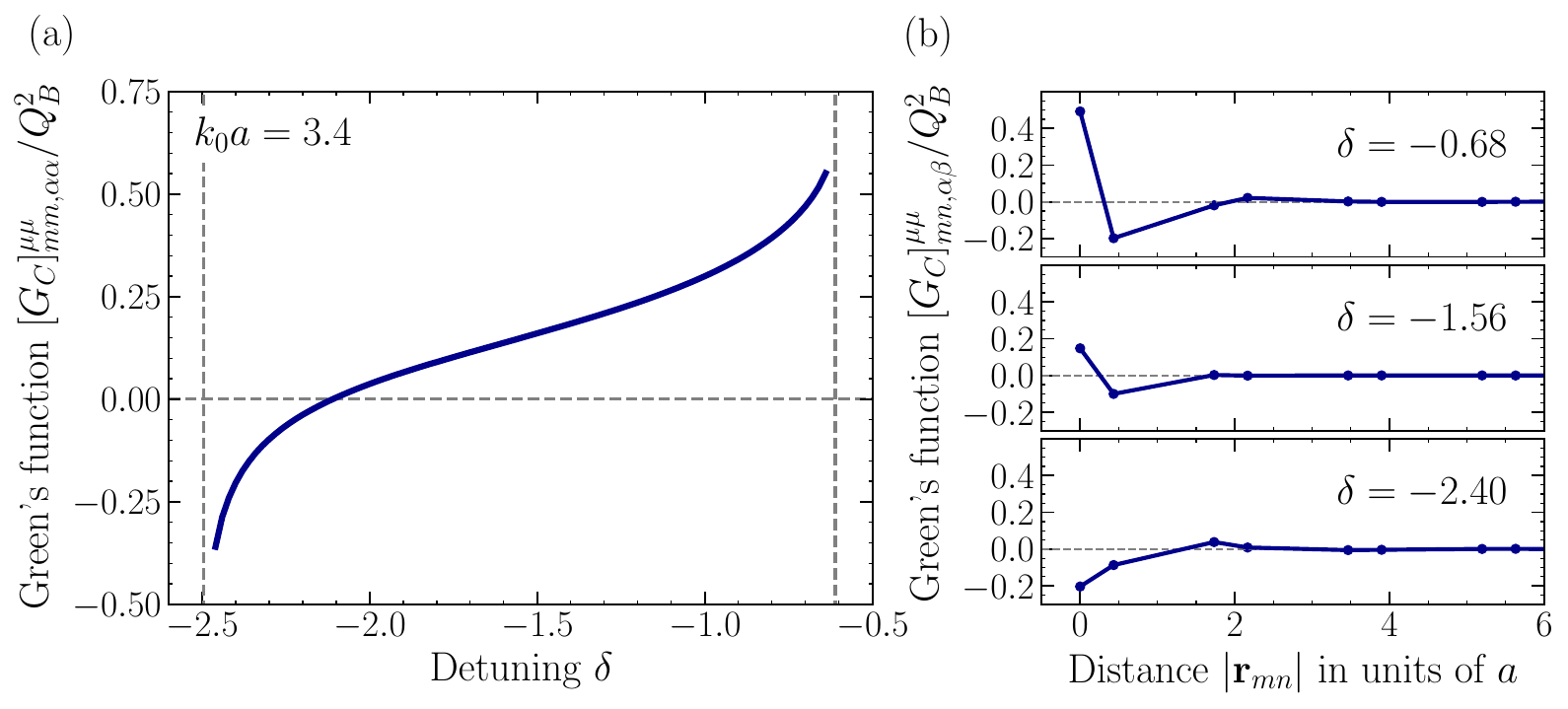}
       \caption{ \label{fig:gf34} Green's function for a diamond crystal of point scatterers with \(k_0a=3.4\), computed using \(10^6\) \(\mathbf{q}\) vectors uniformly sampled in the first Brillouin zone. 
       (a) Dependence of $[G_{\text{C}}]_{mm, \alpha\alpha}^{\mu\mu}$ on the detuning \(\delta\) within the photonic band gap. 
       (b) Dependence of $[G_{\text{C}}]_{mn, \alpha\beta}^{\mu\mu}$ on the distance  $|\vec{r}_{mn}|$ between scatterers in the crystal for three values of detuning, corresponding to frequencies near the high-frequency band edge, the center of the photonic band gap, and near the low-frequency band edge.}
\end{figure*}

Figure~\ref{fig:gf34}(a) shows in greater detail the on-site Green's function for a crystal with \(k_0a = 3.4\) and detunings within the photonic band gap, whose edges are represented by the dashed lines. At this lattice constant, the Green's function is negative up to \(\delta=-2.1\), before becoming positive. 
Let us turn to the distance dependence of the function. 
Figure~\ref{fig:gf34}(b) shows the elements $[G_{\text{C}}]_{mn, \alpha\beta}^{\mu\mu}$ as functions
of the distance \(r_{mn}\) between scatterers along the diagonal direction of the crystal for three representative detunings, corresponding to the low-frequency band edge, the center of the photonic band gap, and the high-frequency band edge. 
These elements of the Green's function decay to zero within the range of two unit cells. The physical interpretation of the magnitude and sign of $[\operatorname{G}_\text{C}]_{mn, \alpha\beta}^{\mu\mu}$, as well as of the detunings and distances at which it vanishes, are discussed in Sec.~\ref{subsec:DefectsInfinite}.


\section{\label{sec:defect}Defect states inside the photonic band gap}

\subsection{\label{subsec:DefectsInfinite}Infinite crystal}
\label{sec:infinite_defect}

After reviewing the properties of ideal defect-free diamond crystals, we now discuss the impact that introducing a defect have on its spectrum. In particular, we are interested in the possibility of creating defect-induced states at frequencies inside the photonic band gap. 
To this end, we replace the scatterer at the position $\vec{r}_d$, with original resonance frequency \(\omega_0\),
by an impurity scatterer with resonance frequency \(\omega_d = \omega_0 + \Delta\omega_0\). The
resonance linewidth of the impurity scatter is assumed to remain unchanged, $\Gamma_d = \Gamma_0$.
The dimensionless detuning corresponding to \(\omega_d\), which quantifies the relative strength of the defect, is \(\delta_d = 2\Delta\omega_0/\Gamma_0\).
The crystal's permittivity is given by 
\begin{equation}
    \epsilon_{d}(\mathbf{r}) = \epsilon(\mathbf{r}) + \Delta\alpha_{\text{B}}\delta(\mathbf{r}-\mathbf{r}_d),
    \label{eq:EpsilonDefect}
\end{equation}
where \(\epsilon(\mathbf{r})\) is the permittivity of the ideal defect-free diamond crystal \eqref{eq:Epsilon} and \(\Delta\alpha_{\text{B}} = \alpha_d - \alpha_{\text{B}}\) is the difference between the bare polarizability of the impurity scatterer and the bare polarizability of all other scatterers in the
host crystal. 
Light propagating in the doped crystal obeys 
\begin{equation}
    -\nabla \times \nabla \times \mathbf{E} +\epsilon(\mathbf{r})\left(\frac{\omega}{c}\right)^2\mathbf{E} = -\Delta\epsilon(\mathbf{r})\left(\frac{\omega}{c}\right)^2\mathbf{E}
    \label{eq:WaveEquationDisordered}
\end{equation}
with \(\Delta\epsilon(\mathbf{r}) = \epsilon_d(\mathbf{r}) - \epsilon(\mathbf{r}) =\Delta\alpha_\mathrm{B}\delta(\vec{r}-\vec{r}_\mathrm{d})\). 
Equation~\eqref{eq:WaveEquationDisordered} is formally rewritten as an integral equation using the crystal's Green's function introduced in Sec.\ \ref{sec:GreensFunction}
\begin{equation}
    \mathbf{E}(\mathbf{r}) = \int d\mathbf{r}' \mathcal{G}_\text{C}(\mathbf{r}, \mathbf{r}', \omega)\left[\Delta\epsilon(\mathbf{r'})\left(\frac{\omega}{c}\right)^2\mathbf{E}(\mathbf{r}')\right].
    \label{eq:FieldDefect}
\end{equation}
Equation \eqref{eq:FieldDefect} admits solutions at frequencies $\omega$ within the photonic band gap of the ideal defect-free crystal provided that the determinantal condition 
\begin{equation}
    \operatorname{det}\left[
    \left( \frac{c}{\omega} \right)^2 \frac{\mathbb{1}_3}{\Delta \alpha_{\text{B}}} + \mathcal{G}_\text{C}(\vec{r}_{d}, \vec{r}_{d}, \omega) \right] = 0
    \label{eq:SingleDefectConditionFunc}
\end{equation}
is obeyed. From \(\omega_0^2=6\pi c^2/\Lambda_\mathrm{T}\alpha(0)\) and the definitions of \(\alpha(0)\) and \(\Lambda_\mathrm{T}\) introduced in Sec.~\ref{subsec:FiniteLattices}, we obtain that variations in the bare polarizability are related to variations in frequency by
\begin{equation}
    \frac{1}{\Delta\alpha_\mathrm{B}} = - \frac{1}{4\alpha_\mathrm{B}^2}\frac{6\pi}{k_0^3}\frac{\Gamma_0}{\Delta\omega_0}\, .  
\end{equation}
Thus, we can write Eq.\ \eqref{eq:SingleDefectConditionFunc} as
\begin{equation}
\operatorname{det}\left[- \frac{\mathbb{1}_3}{4\alpha_\mathrm{B}^2}\frac{6\pi}{k_0^3}\frac{\Gamma_0}{\Delta\omega_0}\left(\frac{c}{\omega}\right)^2 + \mathcal{G}_\text{C}(\vec{r}_{d}, \vec{r}_{d}, \omega) \right] = 0
\end{equation}
or, equivalently, as
\begin{equation}
    \operatorname{det}\left[\frac{\mathbb{1}_3}{\delta_d} + \frac{2}{Q_{\text{B}}^2}\left[\operatorname{G}_\mathrm{C}(\delta)\right]_{mm, \alpha\alpha} \right] = 0.
    \label{eq:SingleDefectCondition}
\end{equation}
Here, $\mathbb{1}_3$ is a $3 \times 3$ identity matrix and $\left[
\operatorname{G}_\mathrm{C}(\delta)\right]_{mm, \alpha\alpha}$ is a $3\times 3$ matrix with elements $\left[\operatorname{G}_\mathrm{C}(\delta)\right]_{mm, \alpha\alpha}^{\mu\nu}$, $\mu,\nu = x, y, z$. 

Equation~\eqref{eq:SingleDefectCondition} is the vector analogue of a scalar equation that conditions the existence of a bound state due to a substitutional impurity in a tight-binding model \cite{economou2006green}. 
The equation admits solutions only for frequencies within the photonic band gap, since \([
\operatorname{G}_\text{C}(\delta)]_{mm,\alpha\alpha}\) diverges for $\delta$ within bands of allowed states.
Equation~\eqref{eq:SingleDefectCondition} is independent of the defect position \(\mathbf{r}_d\), as expected for a translationally invariant infinite crystal. 

\begin{figure}
    \centering
    \includegraphics[width=1\linewidth]{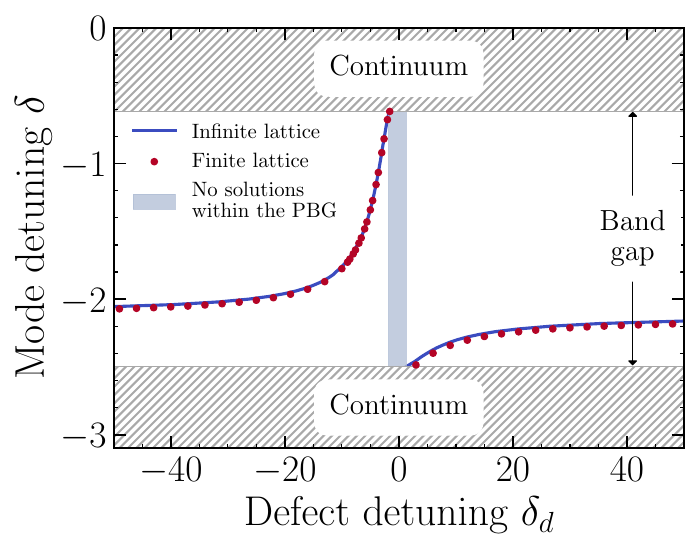}
    \caption{ 
    Detuning $\delta$ of the mode induced within the photonic band gap of a diamond photonic crystal with \(k_0a=3.4\) versus defect detuning $\delta_d$.
    Solid blue lines show results obtained for an infinite crystal from Eq.\ \eqref{eq:SingleDefectCondition}, red dots show results obtained for a crystal of finite size \(k_0 R = 15\), containing \(N=2869\) scatterers.
    The shaded region around \(\delta_d=0\) indicates a range of $\delta_d$ for which Eq.\ \eqref{eq:SingleDefectCondition} admits no solution within band gap.}
    \label{fig:SingleDefectSolutions}
\end{figure}

The solid blue lines in Fig.~\ref{fig:SingleDefectSolutions} show the solutions of Eq.~\eqref{eq:SingleDefectCondition}, which relates the defect detuning \(\delta_d\) to the detuning \(\delta\) of the corresponding defect-induced mode inside the photonic band gap, for a diamond crystal with \(k_0a=3.4\). 
The blue shaded region corresponds to detunings \(\delta_d\) for which Eq.\ \eqref{eq:SingleDefectCondition} has no solution within the gap. 
When the defect is strongly detuned from the host scatterers, \(\delta_d\rightarrow \pm \infty\), the term proportional to \(1/\delta_d\) in Eq.~\eqref{eq:SingleDefectCondition} vanishes. 
In this limit, that corresponds to the creation of a vacancy in the crystal, the condition for creating a state within the photonic band gap becomes
\begin{equation}
    \det\left[\left[
    \operatorname{G}_\mathrm{C}(\delta)\right]_{mm,\alpha\alpha}\right]=0.
    \label{eq:VacancyCondition}
\end{equation}
Since the on-site Green's function \(\left[\operatorname{G}_\mathrm{C}(\delta)\right]_{mm,\alpha\alpha}\) is diagonal and proportional to the identity matrix, it follows that the single-vacancy state is expected to occur at a zero of the Green's function \(\mathcal{G}_\text{C}(\mathbf{r}_d,\mathbf{r}_d, \omega) = 0\). For \(k_0a=3.4\), this vacancy-induced state appears at \(\delta = -2.1\), as shown in Fig.~\ref{fig:SingleDefectSolutions}. 
As discussed in Sec.~ \ref{subsec:InfiniteLattices}, the Green's function for sparse crystals with \(k_0a \geq 4.8\) is strictly positive. 
Therefore, we expect that introducing a vacancy in this regime will not induce a state within the gap. 

Previous studies of partially filled diamond crystals with lattice spacing \(k_0a=2\) and different filling factors revealed a major peak in the DOS around \(\delta \simeq -6.16\), which was attributed to single-vacancy states \cite{antezza2013photonic}. 
Equation\ \eqref{eq:SingleDefectCondition} yields \(\delta \simeq -6.08\) when \(\delta_d \rightarrow - \infty\) and \(\delta \simeq -6.1\) when \(\delta_d \rightarrow \infty\), confirming that this peak is
indeed due to states induced by vacancies.


\subsection{\label{subsec:DefectsFinite}Finite crystal}

Although the model for defects in infinite diamond photonic crystals presented in the previous section establishes the correspondence between the defect scatterer detuning and the induced state within the band gap, experimentally relevant systems have finite sizes. 
Considering the case of scatterers with large quality factors \(Q\), a defect is introduced as a diagonal perturbation of the effective Hamiltonian in Eq.~\eqref{eq:effective_hamiltonian}, yielding 
\begin{equation}
    \mathcal{H}_d = \mathcal{H} + d\mathcal{H}, 
\end{equation}
where \(d\mathcal{H}\) is a \(3N\times3N\) block-diagonal matrix with \(3\times3\) diagonal blocks \((d\mathcal{H})_{mm} = \delta_{md} \delta_d\frac{\Gamma_0}{2}\mathds{1}_3\) for a defect located at the site \(\mathbf{r}_d\). 
Similarly to the unperturbed case, the resonances of the perturbed crystal satisfy $\operatorname{det}\left[\left(\omega_l - i \Gamma_l/2\right)\mathds{1}-\mathcal{H}_d \right]=0$.

\begin{figure}
    \centering
    \includegraphics[width=1\linewidth]{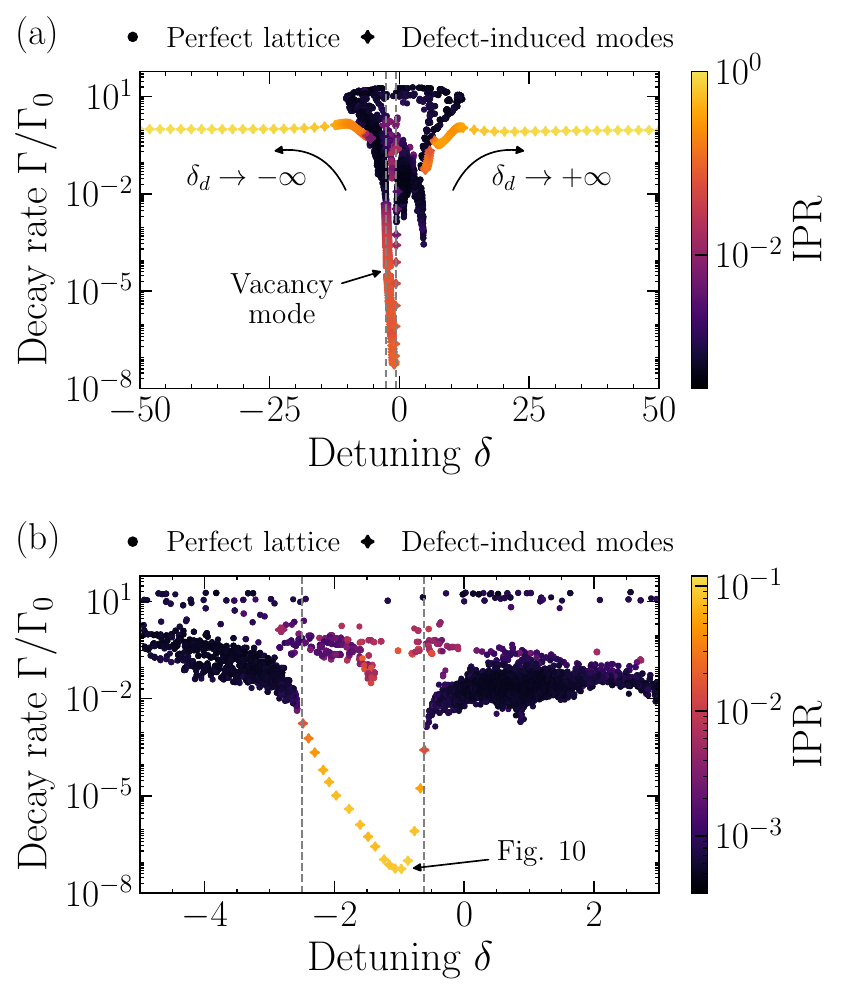}
    \caption{Superposition of resonances of the ideal defect-free diamond crystal of point scatterers shown in Fig.\ \ref{fig:ResonancesDefectlessLattice} with defect-induced resonances inside and outside the photonic band gap for detunings \(\delta_d \in [-50, 50]\). 
    The defect scatterer is placed at the center of a finite crystal of radius \(k_0R=15\) and lattice spacing \(k_0a=3.4\). 
    For clarity, only one representative mode of each triplet is shown for each simulation. The complete spectrum in panel (a) emphasizes the defect-induced resonances that appear outside the band gap, while panel (b) shows a zoom on the region within the band gap. Each defect-induced resonance is obtained from an independent simulation. 
    The color scale shows the IPR of quasinormal modes corresponding to the resonances shown in the figure.}
    \label{fig:ResonancesPerturbedLattice}
\end{figure}

Let us consider the same finite spherically-shaped diamond photonic crystal as in Sec.~\ref{sec:ResonancesInDefectFreeCrystals}. 
When a defect is introduced at \(\vec{r}_d = 0\), the spectrum shown in Fig.~\ref{fig:ResonancesDefectlessLattice} is modified. 
For the vast majority of quasinormal modes this modification is negligible. 
A notable exception consists of six quasinormal modes: three modes emerging within the photonic band gap and three modes appearing outside the gap, all characterized by high values of IPR. 
The three modes within each triplet are degenerate up to numerical precision.
Figure~\ref{fig:ResonancesPerturbedLattice} shows a superposition of the resonances of the ideal defect-free diamond crystal with the defect-states obtained from independent simulations in which a single defect with \(\delta_d \in [-50,50]\) is introduced in the crystal.

Whereas defect-induced states inside photonic band gap are predicted by the infinite-crystal theory of Sec.\ \ref{sec:infinite_defect}, the emergence of additional states outside band gap, and beyond the extent of the spectrum of the defect-free crystal, is a new result
that we obtain only from the calculation for a crystal of finite size.
As shown in Fig.~\ref{fig:ResonancesPerturbedLattice}(a), when the defect scatterer is strongly detuned from the scatterers of the host crystal, such defect-induced modes appear at \(\delta \simeq \delta_d\) and feature decay rates \(\Gamma/\Gamma_0 \simeq 1\). 
They are strongly localized on the defect scatterer (\(\operatorname{IPR}\simeq 1\)) and correspond to single-scatterer excitations that do not couple to the photonic crystal. 
Within the photonic band gap, the behavior is different. When the defect is strongly detuned, the defect-induced modes approaches an asymptotic frequency \(\delta \simeq -2.13\), with decay rates
\(\Gamma/\Gamma_0 \simeq 4\times 10^{-5}\), and converges to the single-vacancy mode discussed in Sec.~\ref{subsec:DefectsInfinite}. 
This asymptotic behavior is not visible in Fig.~\ref{fig:ResonancesPerturbedLattice}(b) because the defect detunings \(\delta_d\) were chosen so as to produce defect-induced modes that uniformly fill the photonic band gap.

When the defect detuning approaches the unperturbed crystal's spectrum, collective effects come into play and the defect-induced modes hybridize with the modes of the photonic crystal. This hybridization delocalizes the modes, decreasing \(\operatorname{IPR}\) to values of the order of $10^{-1}$ and decreasing decay rates to \(\Gamma/\Gamma_0 \sim 10^{-1}\). Further decrease of $\delta_d$ eventually brings the defect-induced modes inside the spectrum of the perfect crystal. 

\begin{figure}
    \centering
    \includegraphics[width=\linewidth]{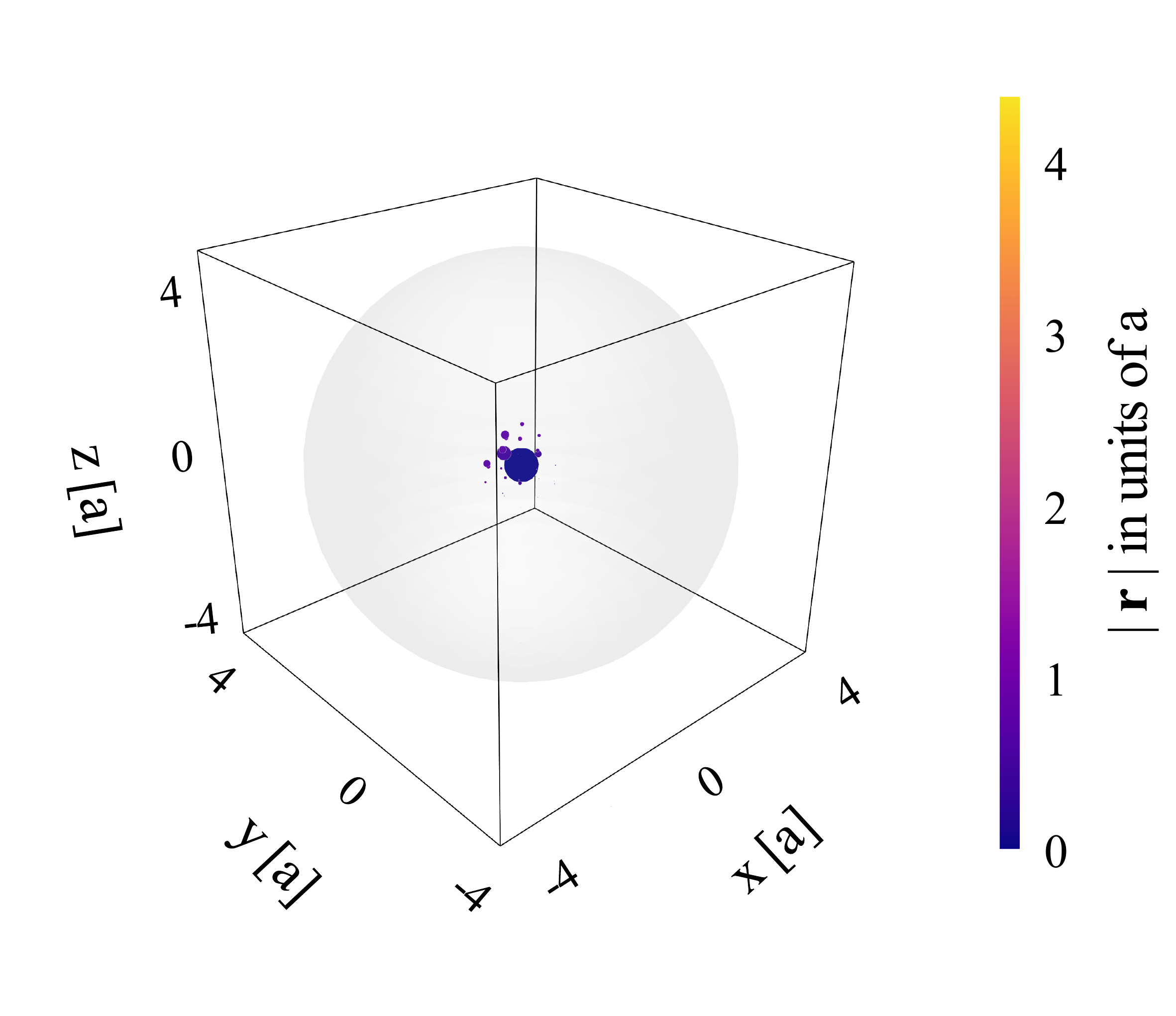}
    \caption{Spatial profile of the quasinormal mode corresponding to the eigenvalue indicated by an arrow in Fig.~\ref{fig:ResonancesPerturbedLattice}(b), in the same representation as for modes in Fig.\ \ref{fig:DefectlessLatticeModes}.
    The mode is induced by a defect with \(\delta_d = -3.2\).
    }
\label{fig:LongestLivedMode}
\end{figure}

Figure~\ref{fig:ResonancesPerturbedLattice}(b) shows a zoom of the spectral region around the gap, displaying a selection of eighteen representative defect-induced modes arising inside the photonic band gap. Detunings of these modes are in good agreement with the predictions of the infinite-crystal theory developed in Sec.\ \ref{sec:infinite_defect}, as we show in Fig.\ \ref{fig:SingleDefectSolutions}. In addition our calculation also yields decay rates $\Gamma$ of the modes, which vanish in the case of the infinite crystal. 
As is apparent from Fig.~\ref{fig:ResonancesPerturbedLattice}(b), decay rates of defect modes inside the photonic band gap are significantly lower than the typical decay rates of the modes of the ideal defect-free crystal. They reach a minimum value of
\(\Gamma/\Gamma_0 = 5.49 \times 10^{-8}\) for the mode at \(\delta = -1.02\). The spatial structure of this longest-lived defect mode is shown in Fig.~\ref{fig:LongestLivedMode}.
The mode is strongly localized in the vicinity of the defect scatterer, which is also the case for all other defect modes arising inside the band gap for other values of $\delta$, as witnessed by their large IPR in Fig. \ref{fig:ResonancesPerturbedLattice}(b).

Modes near the high-frequency band edge feature longer lifetimes and higher values of \(\operatorname{IPR}\) compared to those close to the low-frequency edge. This asymmetry, which is clearly seen in Fig.~\ref{fig:ResonancesPerturbedLattice}(b), is attributed to the residual DOS within the gap due to the finite crystal size, which is smaller on the high-frequency side, thereby favoring the formation of spatially localized states. This interpretation is supported by a comparison of results obtained for crystals of different shapes.
As we show in Fig.~\ref{fig:DOSFinite}, the DOS of a crystal having the shape of a cube exhibits a wider plateau on the high-frequency side of the gap than the DOS of the spherical sample. Repeating calculations of this section for a crystal of cubic shape (results not shown), we find that defect-induced modes with detunings along this plateau tend to display lower decay rates and higher \(\operatorname{IPRs}\) than their counterparts in the crystal of spherical shape.

\begin{figure*}
    \centering
    \includegraphics[width=0.9\linewidth]{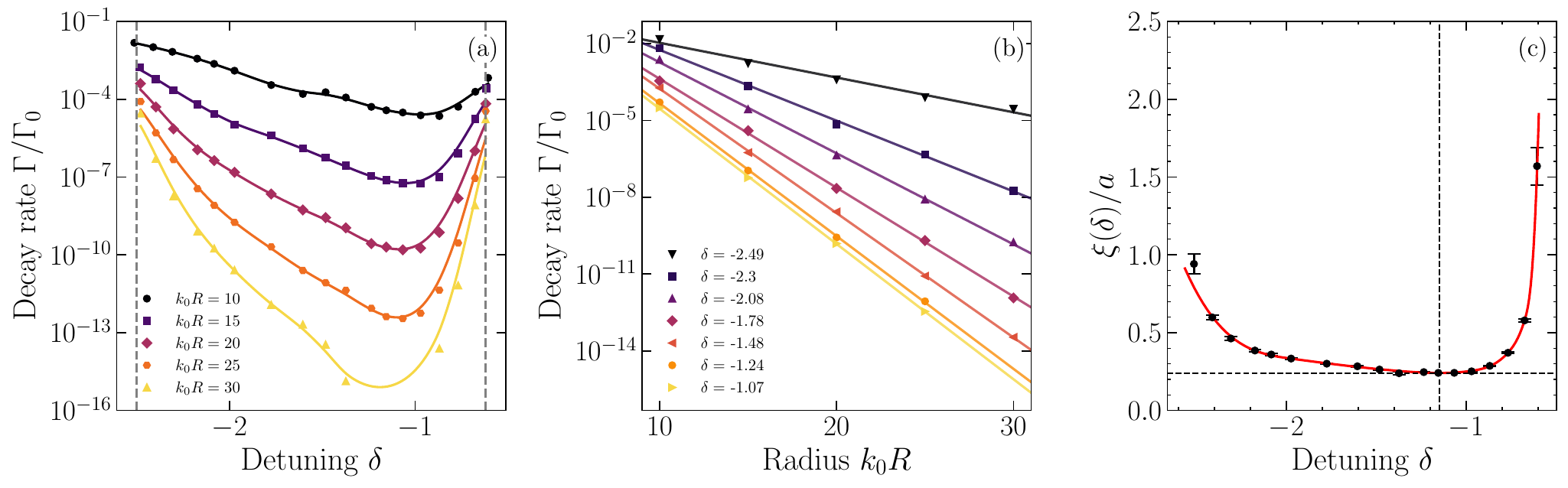}
    \caption{
    (a) Normalized decay rates of defect-induced quasinormal modes as functions of detuning inside the photonic band gap delimited by vertical dashed lines, for different crystal  sizes \(k_0R\) (symbols). 
    The defect is placed at the center of the diamond crystal with \(k_0a=3.4\). The radii \(k_0R = 10\), 15, 20, 25 and 30 correspond to \(N = 849\), 2869, 6851, 13331 and 22929, respectively. 
    Defect detunings
    span \(\delta_d \in [-30,30]\).
    (b) Scaling of decay rates with crystal size \(k_0R\) at seven 
    representative detunings within the photonic band gap (symbols). Solid lines
    show exponential fits of the form \(\Gamma/\Gamma_0 = \gamma(\delta) e^{-R/\xi(\delta)}\). 
    (c) The best-fit localization length \(\xi(\delta)\) in units of lattice spacing $a$ and as a function of  detuning \(\delta\) (symbols).
    Solid lines in panels (a) and (c) are cubic smoothing splines used only as guides to the eye.
    }
    \label{fig:scaling}
\end{figure*}

Figure~\ref{fig:scaling}(a) shows the behavior of the decay rates as the radius of the finite crystal is increased from \(k_0R=10\) up to \(k_0R=30\). The longest-lived mode always occurs in the vicinity of \(\delta \simeq -1.05\), with decay rates decreasing from \(\Gamma/\Gamma_0 = 2.25 \times 10^{-5}\) for \(k_0R = 10\) to \(\Gamma/\Gamma_0 = 3.48 \times 10^{-13}\) for \(k_0R=25\). For \(k_0R = 30\), four points corresponding to the modes with the longest lifetimes can not be retrieved because their decay rates fall below the numerical precision of our calculations, i.e., \(\Gamma/\Gamma_0 < 10^{-16}\). 
Figure~\ref{fig:scaling}(b) demonstrates that the decay rates
scale exponentially with crystal size, with
$\Gamma/\Gamma_0 \propto \exp(-R/\xi)$. The corresponding localization length $\xi$ depends on mode detuning but remains always of the order of crystal lattice constant $a$, as is apparent in Fig.~\ref{fig:scaling}(c). 
This confirms the strongly localized character of quasinormal modes induced by isolated defects in an otherwise ideal defect-free photonic crystal.

\section{Conclusions and Outlook}\label{sec:conclusion}

We have investigated the emergence of defect-induced modes within the photonic band gap of finite and infinite three-dimensional photonic crystals composed of point-like resonant scatterers arranged in a diamond lattice. The defect is introduced by changing the resonance frequency of one of the scatterers. In the infinite-crystal limit, we establish a correspondence between the defect and mode detunings via the determinantal condition in Eq.~\eqref{eq:SingleDefectCondition}, which involves the crystal's Green’s function calculated in Sec.\ \ref{subsec:InfiniteLattices}. While the infinite-crystal model allows us to predict the frequencies of defect-induced modes, it fails to capture finite-size effects that are unavoidable in experiments.
To address this deficiency, we turn to a model of a photonic crystal of finite size that allows us to discover that a single defect gives rise to six localized quaismodes: three degenerate modes inside and three degenerate modes outside the photonic band gap.
Eigenfreqencies of the modes arising inside the band gap agree very well with the infinite-crystal calculation, whereas their lifetimes increase exponentially with the crystal size, which is consistent with their strong spatial localization in the vicinity of the defect.
The modes emerging outside the band gap are clearly identified only when the defect is sufficiently strong to bring their frequencies beyond the spectrum of the ideal defect-free crystal. 
They correspond to the resonance of the defect scatterer decoupled from the rest of the system and thus have the frequency and the lifetime of the defect scatterer in the free space.

An experimental situation for which calculations presented in this paper are relevant may correspond to a photonic crystal with many identical defects at sufficiently distant locations. Provided that different defects do not interact (single scattering approximation), they should introduce identical defect states inside the band gap and could be detected by measuring the time-resolved transmission $\mathcal{T}(t)$ of a short optical pulse through the crystal. Localized defect states discussed in this work should give rise to long-time tails of $\mathcal{T}(t) \propto \exp(-\Gamma t)$ that decay orders of magnitude slower than the typical decay of $\mathcal{T}(t)$ through an ideal, defect-free crystal. 
Alternatively, the frequency-resolved continuous-wave transmission $\mathcal{T}(\omega)$ should feature a narrow transmission band of width $\Gamma$ around the frequency of the defect-induced state inside the photonic band gap.

As an outlook, it will be interesting to extend the theoretical framework established in this paper to the physical situation of many coupled defects to study the multiple scattering of light between defects, including hopping known as ``Cartesian light''~\cite{hack2019cartesian}, and explore the possibility of 3D Anderson localization of light in such a tight-binding setting. 
Indeed, Anderson localization has been recently predicted to occur in a dense ensemble of identical defects in a transparent material, when a strong external magnetic field lifts the degeneracy of the three polarization states \cite{skipetrov2025prb}. 
Using defect states inside a photonic band gap instead of a transparent material has an obvious advantage of forbidding free propagation and ``forcing'' a photon to hop between nearest-neighbor defects only, which should favor localization. In addition, the theoretical study of such a system may be substantially facilitated by the fact that we now know the Green's function of the ideal photonic crystal [Eq.\ \eqref{gcfinalq} and Figs.\ \ref{fig:green_function} and \ref{fig:gf34}], which can be used instead of the free-space Green's function in numerical studies. As a result, the computational cost of solving the problem may be greatly reduced by reducing the size of the matrix to diagonalize from  \(3N\times3N\) to \(3N_\text{d}\times3N_\text{d}\), where $N$ and $N_\text{d} \ll N$ are the total number of scatterers and the number of defects, respectively.

Finally, even if the present work deals with a particular crystal of point-like scatterers, we believe that our main
conclusions as well as the experimental signatures of defect states should be similar for any photonic crystal with a band gap and any type of local defect. Thus, our results highlight the potential of engineered defects in three-dimensional photonic band gap crystals for the realization of long-lived, spatially localized states. Our findings may open a way to the controlled design of defect-based photonic modes for applications in quantum light–matter interfaces.

\section{\label{sec:acknowledgments}Acknowledgments}
We commemorate our co-author Bart van Tiggelen, who passed untimely, and who was always an untiring and enthusiastic proponent of this research. 
WLV thanks the CNRS for supporting his stay as an invited research professor the LPMMC.
WLV and AL acknowledge support by the Dutch Research Council NWO-TTW Perspectief program P21-20 ‘Optical coherence; optimal delivery and positioning’ (OPTIC) in collaboration with TUE, TUD, and ARCNL, and with industrial partners Anteryon, ASML, Demcon, JMO, Signify, and TNO.

\appendix

\section{\label{ap:hamiltonian}Effective Hamiltonian for an infinite diamond crystal}

In this appendix, we derive the expression for the effective Hamiltonian of an infinite diamond crystal presented in Sec.~\ref{subsec:InfiniteLattices}.

From the set of coupled equations \eqref{eq:coupled}, the field exciting the point scatterer \(A\) in the \(m^\text{th}\) unit cell of a diamond crystal can be written as the superposition of the fields scattered by all other scatterers occupying \(A\) sites and the fields scattered by all the scatterers at
\(B\) sites, and reads
\begin{multline}
        \mathbf{E}(\vec{R}_m - \mathbf{d}/2) = \mathbf{E}_0(\vec{R}_m - \mathbf{d}/2) \\ - k^2\alpha(\omega) \sum_{n\neq m} \mathcal{G}_0(\vec{R}_m - \mathbf{d}/2, \vec{R}_n - \mathbf{d}/2)\mathbf{E}(\mathbf{R}_n - \mathbf{d}/2) \\ - k^2\alpha(\omega) \sum_{n} \mathcal{G}_0(\mathbf{R}_m - \mathbf{d}/2, \mathbf{R}_n + \mathbf{d}/2)\mathbf{E}(\mathbf{R}_n + \mathbf{d}/2),
        \label{eq:FieldA}
\end{multline}
where the sums are performed considering an infinite number of scatterers in the crystal, \(N\to\infty\). Following the same logic, the field exciting the scatterer \(B\) of the same unit cell is given by
\begin{multline}
        \mathbf{E}(\mathbf{R}_m + \mathbf{d}/2) = \mathbf{E}_0(\mathbf{R}_m + \mathbf{d}/2)
        \\
- k^2\alpha(\omega) \sum_{n=1} \mathcal{G}_0(\mathbf{R}_m + \mathbf{d}/2, \mathbf{R}_n - \mathbf{d}/2)\mathbf{E}(\mathbf{R}_n - \mathbf{d}/2)    
\\
- k^2\alpha(\omega) \sum_{n \neq m}\mathcal{G}_0(\mathbf{R}_m + \mathbf{d}/2, \mathbf{R}_n + \mathbf{d}/2)\mathbf{E}(\mathbf{R}_n + \mathbf{d}/2).
        \label{eq:FieldB}
\end{multline}
According to Bloch's theorem, the field exciting each scatterer in the infinite crystal can be written as a
sum of Bloch modes
\begin{equation}
    \vec{E}(\vec{r}_m) = \vec{u}_\vec{q}^{(\alpha)}(\vec{r}_m)e^{i\vec{q}\cdot\vec{r}_m},
    \label{eq:BlochMode}
\end{equation}
with \(\alpha=A, B\) and \(\vec{u}_\vec{q}^{(\alpha)}\) a function with the periodicity of the underlying fcc structure generating the diamond crystal, such that 
\begin{equation}
    \vec{u}_\vec{q}^{(\alpha)}(\vec{r}+\vec{R}) = \vec{u}_\vec{q}^{(\alpha)}(\vec{r}),
    \label{eq:BlochModePeriodicity}
\end{equation}
for all \(\vec{R}\) in the underlying fcc lattice. Therefore, introducing the form \eqref{eq:BlochMode} in Eq.~\eqref{eq:FieldA} yields 
\begin{multline}
   \vec{u}_\vec{q}^{(A)}(\vec{R}_m -\vec{d}/2)e^{-i\vec{q}\cdot\frac{\vec{d}}{2}} + \\
   k^2\alpha(\omega)\sum_{n\neq m}\mathcal{G}_0(\vec{R}_{mn})e^{-i\vec{q}\cdot\vec{R}_{mn}}\vec{u}_\vec{q}^{(A)}(\vec{R}_n -\vec{d}/2)e^{-i\vec{q}\cdot\frac{\vec{d}}{2}} + \\
   k^2\alpha(\omega)\sum_{n}\mathcal{G}_0(\vec{R}_{mn}-\vec{d})e^{-i\vec{q}\cdot(\vec{R}_{mn})}\vec{u}_\vec{q}^{(B)}(\vec{R}_n+\vec{d}/2)e^{i\vec{q}\cdot\frac{\vec{d}}{2}} \\ = 0,
\end{multline}
in the absence of incident fields, \(\vec{E}_0 = 0\), and \(\vec{R}_{mn} = \vec{R}_m - \vec{R}_n\). Similarly, from Eq.~\eqref{eq:FieldB} we have
\begin{multline}
   \vec{u}_\vec{q}^{(B)}(\vec{R}_m +\vec{d}/2)e^{i\vec{q}\cdot\frac{\vec{d}}{2}} + \\
   k^2\alpha(\omega)\sum_{n}\mathcal{G}_0(\vec{R}_{mn}+\vec{d})e^{-i\vec{q}\cdot(\vec{R}_{mn})}\vec{u}_\vec{q}^{(A)}(\vec{R}_n -\vec{d}/2)e^{-i\vec{q}\cdot\frac{\vec{d}}{2}} + \\
   k^2\alpha(\omega)\sum_{n\neq m}\mathcal{G}_0(\vec{R}_{mn})e^{-i\vec{q}\cdot\vec{R}_{mn}}\vec{u}_\vec{q}^{(B)}(\vec{R}_n+\vec{d}/2)e^{i\vec{q}\cdot\frac{\vec{d}}{2}} \\ = 0.
\end{multline}
Using the periodicity of Bloch functions \eqref{eq:BlochModePeriodicity}, we have
\begin{equation}
    \vec{u}_\vec{q}(\vec{R})\left[\mathds{1}+k^2\alpha(\omega)\mathds{G}(\vec{q},\omega)\right] = 0
    \label{ap:eq:ResonanceEquation}
\end{equation}
with the vector
\begin{equation}
    \vec{u}_\vec{q}(\vec{R}) = \left[\begin{matrix}
        \vec{u}_\vec{q}^{(A)}(\vec{R}-\vec{d}/2)e^{-i\vec{q}\cdot\frac{\vec{d}}{2}}\\ \vec{u}_\vec{q}^{(B)}(\vec{R}+\vec{d}/2)e^{i\vec{q}\cdot\frac{\vec{d}}{2}}
    \end{matrix}\right],
\end{equation}
and where
\begin{equation}
    \mathds{G}(\vec{q},\omega) = \left(\begin{matrix}
        \mathds{G}_{AA} & \mathds{G}_{AB}\\ 
        \mathds{G}_{BA} & \mathds{G}_{BB}
    \end{matrix}\right)
\end{equation}
is a \(6\times6\) matrix with elements given by the four sums
\begin{equation}
    \mathds{G}_{AA} = \sum_{\vec{R}\neq 0}\mathcal{G}_0(\vec{R})e^{-i\vec{q}\cdot\vec{R}},
    \label{ap:eq:G_AA}
\end{equation}
\begin{equation}
    \mathds{G}_{AB} = \sum_{\vec{R}}\mathcal{G}_0(\vec{R}-\vec{d})e^{-i\vec{q}\cdot\vec{R}},
    \label{ap:eq:G_AB}
\end{equation}
\begin{equation}
    \mathds{G}_{BB} = \sum_{\vec{R}\neq 0}\mathcal{G}_0(\vec{R})e^{-i\vec{q}\cdot\vec{R}},
    \label{ap:eq:G_BB}
\end{equation}
\begin{equation}
    \mathds{G}_{BA} = \sum_{\vec{R}}\mathcal{G}_0(\vec{R}+\vec{d})e^{-i\vec{q}\cdot\vec{R}}.
    \label{ap:eq:G_BA}
\end{equation}
The sums over the direct lattice present in Eqs.~\eqref{ap:eq:G_AA}-\eqref{ap:eq:G_BA} can be transformed into
sums over the reciprocal lattice by using Poisson's summation formula
\begin{equation}
    \frac{1}{\Omega}\sum_{\vec{Q}}e^{i\vec{Q}\cdot\vec{r}} = \sum_{\vec{R}}\delta(\vec{r}-\vec{R}),
    \label{ap:eq:PoissonSummationFormula}
\end{equation}
yielding 
\begin{equation}
    \mathds{G}_{AA} = \mathds{G}_{BB} = \frac{1}{\Omega}\sum_{\vec{Q}}\hat{\mathcal{G}}_0(\vec{Q}-\vec{q}, \omega) - \mathcal{G}_0(\vec{r},\vec{r},\omega),
    \label{ap:eq:G_AA_Q}
\end{equation}
\begin{equation}
    \mathds{G}_{AB} = \frac{1}{\Omega}\sum_{\vec{Q}}\hat{\mathcal{G}}_0(\vec{Q}-\vec{q},\omega)e^{i(\vec{Q}-\vec{q})\cdot\vec{d}},
    \label{ap:eq:G_AB_Q}
\end{equation}
\begin{equation}
    \mathds{G}_{BA} = \frac{1}{\Omega}\sum_{\vec{Q}}\hat{\mathcal{G}}_0(\vec{Q}-\vec{q},\omega)e^{-i(\vec{Q}-\vec{q})\cdot\vec{d}}.
    \label{ap:eq:G_BA_Q}
\end{equation}
Solutions to Eq.~\eqref{ap:eq:ResonanceEquation} exist for frequencies \(\omega\) obeying 
\begin{equation}
    \det\left[\mathds{1}+ k^2\alpha(\omega)\mathds{G}(\vec{q},\omega)\right]=0.
    \label{ap:eq:DetCondition}
\end{equation}
Following the same reasoning as presented in Sec.~\ref{subsec:FiniteLattices}, we approximate \(\mathds{G}(\vec{q},\omega) \simeq \mathds{G}(\vec{q} ,\omega_0) \) in Eq.~\eqref{ap:eq:DetCondition} and use the form \eqref{eq:polarizability1} for the polarizability \(\alpha(\omega)\), yielding
\begin{equation}
    \det\left[\left(\omega_\vec{q}-i\frac{\Gamma_\vec{q}}{2}\right)\mathds{1} - \mathcal{H}(
    \vec{q})\right]=0
    \label{ap:eq:DetConditionH}
\end{equation}
with the effective Hamiltonian
\begin{equation}
    \mathcal{H}(
    \vec{q}
    ) = \left(\omega_0 - i \frac{\Gamma_0}{2}\right)\mathds{1} - \frac{\Gamma_0}{2}\tilde{\mathds{G}}(\vec{q},\omega_0),
    \label{ap:eq:HamiltonianInfiniteCrystal}
\end{equation}
where \(\tilde{\mathds{G}}(\vec{q},\omega)= -\frac{6\pi}{k}\mathds{G}(\vec{q},\omega)\). The latter relation between the matrices \(\tilde{\mathds{G}}\) and \(\mathds{G}\), when applied to the sums in Eqs.~\eqref{ap:eq:G_AA_Q}--\eqref{ap:eq:G_BA_Q}, allows to recover the expression for the blocks \(\tilde{\mathds{G}}_{\alpha\beta}\) in Eq.~\eqref{eq:G_alpha_beta}.


\section{\label{ap:green}Green's function in a crystal with diamond structure}
From Eqs.~\eqref{eq:WaveEquation} and \eqref{eq:Epsilon}, we obtain that the field propagating in a diamond crystal satisfy
\begin{equation}
    \mathbf{E}(\mathbf{r})
    =
    \mathbf{E}_0(\mathbf{r})
    -
    \alpha_\mathrm{B}
    \left(\frac{\omega}{c}\right)^2
    \sum_{n=1}^{N}
    \mathcal{G}_0(\mathbf{r}-\mathbf{r}_n)
    \mathbf{E}(\mathbf{r}_n).
\end{equation}
Similarly, the diamond crystal's Green's function can be written as
\begin{equation}
\begin{aligned}
    \mathcal{G}_\text{C}(\mathbf{r},\mathbf{r}',\omega) &= \mathcal{G}_0(\mathbf{r}-\mathbf{r}',\omega) \\
    &- k^2\alpha_{\text{B}}\sum_{j = 1}^N\mathcal{G}_0(\mathbf{r}-\mathbf{r}_j,\omega)\mathcal{G}_\text{C}(\mathbf{r}_j,\mathbf{r}',\omega). 
    \label{ap:eq:CoupledCrystalGreen}
\end{aligned}
\end{equation}
By defining the matrices \((\operatorname{G}_0)_{mn}=-\frac{6\pi}{k}\mathcal{G}_0(\vec{r}_m,\vec{r}_n)\) and \((\operatorname{G}_\text{C})_{mn}=-\frac{6\pi}{k}\mathcal{G}_\text{C}(\vec{r}_m,\vec{r}_n),\  \forall\, m,n\), one can write Eq.~\eqref{ap:eq:CoupledCrystalGreen} as
\begin{equation}
    \operatorname{G}_\mathrm{C}
    =
    \operatorname{G}_0
    +
    \frac{\alpha_\mathrm{B}k^3}{6\pi}
    \operatorname{G}_0 \operatorname{G}_\mathrm{C}
    \label{ap:eq:CrystalsGreensFunctionDysonOperator}
\end{equation}
By iterating over Eq. \eqref{ap:eq:CrystalsGreensFunctionDysonOperator}, one obtains Eq.\eqref{eq:DysonEquation} of Sec.~\ref{sec:GreensFunction}, namely
\begin{equation}
    \operatorname{G}_\mathrm{C}
    =
    \operatorname{G}_0
    -
    \frac{k}{6\pi}
    \operatorname{G}_0 \operatorname{T} \operatorname{G}_0,
    \label{ap:eq:Dyson}
\end{equation}
with the crystal's \(\operatorname{T}\) matrix defined as
\begin{equation}
    \operatorname{T}
    = 
    -\sum_{n=1}^\infty\left[k^2\alpha_\text{B}\right]^n\left[\left(\frac{k}{6\pi}\right)\operatorname{G}_0\right]^{n-1},
\end{equation}
resulting in
\begin{equation}
    \operatorname{T}
    = 
    -
    \frac{
        \mathds{1}
        }{
        \frac{\mathds{1}}{k^2\alpha_\text{B}}
        -
        \frac{k}{6\pi}\operatorname{G}_0}.
        \label{ap:eq:TMatrixGzero}
\end{equation}
By recalling the definitions of the matrices \(\operatorname{G}_0\) and \(\tilde{\mathds{G}}_0\), we obtain the relation
\begin{equation}
    (\operatorname{G}_0)_{mn} = (\tilde{\mathds{G}}_0)_{mn} -  \delta_{mn}\frac{6\pi}{k}\mathcal{G}_0(\vec{r}_m,\vec{r}_n),
    \label{ap:eq:MatrixGDefinition}
\end{equation}
which, when introduced in Eq.~\eqref{ap:eq:TMatrixGzero}, yields
\begin{equation}
    \operatorname{T}
    =
    -
    \frac{
    \mathds{1}
    }{
    \left(\frac{1}{k^2\alpha_\text{B}}+\mathcal{G}_0(\mathbf{r},\mathbf{r})\right)\mathds{1}
     - \frac{k}{6\pi}\tilde{\mathds{G}}_0
    }.
    \label{eq:TMatrixWithBothGMatrices}
\end{equation}
From the regularization procedure introduced in Sec.~\ref{subsec:FiniteLattices}, we obtain 
\begin{equation}
    \frac{1}{\alpha_B}
    \left(\frac{c}{\omega}\right)^2
    +
    \mathcal{G}_0(\mathbf{r},\mathbf{r})
    =
    \frac{1}{k^2\alpha(\omega)},
    \label{ap:eq:RelationAlphasAndG}
\end{equation}
which allows to write the \(\operatorname{T}\)-matrix in Eq.~\eqref{eq:TMatrixWithBothGMatrices} as
\begin{equation}
    \operatorname{T}
    =
    -
    \frac{
    \mathds{1}
    }{
    \frac{1}{k^2\alpha(\omega)}
    -
    \frac{k}{6\pi}\tilde{\mathds{G}}_0
    } = 
    -
    \frac{
    \mathds{1}
    }{
    \frac{1}{k^2\alpha(\omega)}
    +
    \mathds{G}_0
    }  \ .
    \label{eq:TMatrixCrystal}
\end{equation}
From Eq.~\eqref{ap:eq:MatrixGDefinition}, an eigenvector \(\psi_l\) of \(\tilde{\mathds{G}}_0\), associated with the eigenvalue \(\Lambda_l\), is also eigenvector of \(\operatorname{G}_0\) such that
\begin{equation}
    \operatorname{G}_0\psi_l = \left[\Lambda_l-\frac{6\pi}{k}\mathcal{G}_0(\vec{r},\vec{r})\right]\psi_l.
\end{equation}
Following from Eq.~\eqref{ap:eq:Dyson}
\begin{equation}
    \operatorname{G}_\text{C}\psi_l = \zeta_l\psi_l,
\end{equation}
with
\begin{equation}
    \zeta_l = \left[\Lambda_l-\frac{6\pi}{k}\mathcal{G}_0(\vec{r},\vec{r})\right] + \frac{\frac{k}{6\pi}\left[\Lambda_l-\frac{6\pi}{k}\mathcal{G}_0(\vec{r},\vec{r})\right]^2}{\frac{1}{k^2\alpha(\omega)}-\frac{k}{6\pi}\Lambda_l}.
    \label{ap:eq:Zeta1}
\end{equation}
Equation~\eqref{ap:eq:Zeta1} can be simplified to
\begin{equation}
    \zeta_l = \frac{\left[\Lambda_l-\frac{6\pi}{k}\mathcal{G}_0(\vec{r},\vec{r})\right]\left[\frac{1}{k^2\alpha(\omega)}-\mathcal{G}_0(\vec{r},\vec{r})\right]}{\frac{1}{k^2\alpha(\omega)}-\frac{k}{6\pi}\Lambda_l}.
    \label{ap:eq:Zeta2}
\end{equation}
Assuming \(Q_\text{B}=\frac{6\pi}{k^3\alpha_\text{B}}\) and using Eq.~\eqref{ap:eq:RelationAlphasAndG} to eliminate \(\mathcal{G}_0(\vec{r},\vec{r})\) from Eq.~\eqref{ap:eq:Zeta2}, we obtain
\begin{equation}
    \zeta_l = Q_\text{B}\frac{\Lambda_l - \frac{6\pi}{k^3\alpha(\omega)}+Q_\text{B}}{\frac{6\pi}{k}\frac{1}{k^2\alpha(\omega)}-\Lambda_l}.
    \label{ap:eq:Zeta3}
\end{equation}
For \(|\omega-\omega_0| \ll \omega_0\), the polarizability takes the form of Eq.~\eqref{eq:polarizability1}, therefore
\begin{equation}
    \zeta_l \simeq  - Q_\text{B}\frac{Q_\text{B}+\Lambda_l+i}{\delta+\Lambda_l+i}.
    \label{ap:eq:Zeta4}
\end{equation}
For an infinite crystal, \(\operatorname{Im}\Lambda_l = -1\)
and $|\operatorname{Re}\Lambda_l| \ll Q_{\text{B}}$,
thus
\begin{equation}
    \zeta_l \simeq  - \frac{Q_\text{B}^2}{\delta +\operatorname{Re}{\Lambda_l}}.
    \label{ap:eq:Zeta5}
\end{equation}
\\

\hfill


\bibliographystyle{apsrev4-2}
\bibliography{refs_defect_states}


\end{document}